\documentclass[%
aps,
prd,
amsmath,
amssymb,
twocolumn
]{revtex4-2}

\usepackage[pdftex]{graphicx}
\usepackage{dcolumn}
\usepackage{bm}

\newcommand{\aap}{Astron.\ Astrophys.}
\newcommand{\mnras}{Mon.\ Not.\ R.\ Astron.\ Soc.}
\newcommand{\physrep}{Phys.\ Rep.}

\newcommand{\apjl}{Astrophys.\ J.\ Lett.}

\begin{document}

\title{Constraining shear modulus of polycrystalline neutron star crust: Hashin-Shtrikman variational approach}

\author{Nikita A.\ Zemlyakov}
\email{zemlyakov@mail.ioffe.ru}
\author{Andrey I.\ Chugunov}
\email{andr.astro@mail.ioffe.ru}
\affiliation{Ioffe Institute, Politekhnicheskaya 26, St. Petersburg 194021, Russia
}%

\date{\today}

\begin{abstract}
The elastic properties of the neutron star crust are thought to play a crucial role in various phenomena of neutron stars (glitches, oscillations, gravitational wave emission) and should be described quantitatively to model these phenomena.
The fundamental problem of this description is associated with the polycrystalline nature of the crust: similar to terrestrial materials, the elastic moduli, strictly speaking,
depend on the shape and orientation of crystallites, but for the crust, they are unknown. As a result, some assumptions are generally required to predict the elastic properties or constrain their possible range.
In this paper, we follow the commonly
believed assumption that the crust is (locally) isotropic, which allows us to describe elastic properties by two (effective) parameters: bulk and shear moduli. The bulk modulus is well determined by the Voigt-Reuss bounds, and we constrain the shear modulus by applying,
for the first time in astrophysics of compact stars,
the variational Hashin-Shtrikman approach, based on the additional assumption that there are no correlations in the orientation of crystallites.
We analyse the Hashin-Shtrikman bounds for the one-component crust taking into account the electron screening and the motion of the nuclei, and for two-component static crystals. In particular, we demonstrate that within applied assumptions the effective shear modulus should be lower than the Voigt estimate, typically applied in the astrophysical literature. 
\end{abstract}

\maketitle

\section{Introduction}
The neutron star crust is believed to consist of fully ionized atoms (nuclei) on the background of a strongly degenerate gas of (relativistic) electrons (along with unbound neutrons at densities $\rho>\rho_\mathrm{d} \approx 4 \cdot 10^{11}$ g/cm$^3$) \cite{hpy07,ch08}. According to many simulations, performed by different approaches \cite{ oih93,jc96,pc00_EOS_Solid,mc10_crystallization, ch17, Caplan_etal18, Fantina_ea20_outer, Carreau_ea20_inner,Arnold_etal_25}, the nuclei should be crystallized (i.e., form a lattice) at the typical neutron star temperatures, and thus the crust should behave as a solid in a mechanical sense. Thereby the elastic properties of the crust are essential for the interpretation of neutron star observations. 

The crust density increases by several orders of magnitude as one moves into deeper layers; additionally, the composition also changes \cite{hpy07,ch08}. Thus, a microphysical theory should be able to provide quantitative predictions of the elastic properties for the given density and composition.
For this aim, the elastic properties of the crust were well analysed in the approximation of point-like nuclei on a homogeneous background of electrons (the Coulomb crystal approximation); typically, it was also assumed that all nuclei are of the same type (one-component composition). 
Under these assumptions, it is thermodynamically favourable for the nuclei to form a body-centred cubic (bcc) lattice \cite{hfd97, Baiko02, Kozhberov18_Yuk}; however, the electron screening can favour a face-centred cubic lattice for some parameters \cite{Baiko02,kp21}. In any case, the lattice has cubic symmetry and its elastic properties can be described by three elastic coefficients $c_{11}, c_{12}$ and $c_{44}$ \cite{ll_elast}. To date, these coefficients have been reasonably well calculated, taking into account the electron screening \cite{hh08, hh12, Baiko12, Baiko15, Kozhberov22} and the motion of the nuclei (including thermal \cite{oi90, Strohmayer_etal91, hh08, Baiko11, Baiko12, hh12} and quantum \cite{Baiko11, Baiko12} effects). 
The elastic properties of two-component cubic symmetry lattices were also analysed \cite{IH_binary_cryst,Kozhberov19_binary}.

However, it is generally believed that the crust is polycrystalline, i.e. it consists of sufficiently small crystallites (grains).
As a result, in almost all astrophysical applications, the crust is considered as isotropic elastic material. 
Strictly speaking, it is the assumption that can be violated, e.g., due to epitaxial growth of the crust \cite{Baiko24_a, Baiko24_b}.
As shown in \cite{mh24_a, mh25}, even a small anisotropy of the elastic properties of the crust can affect the observable phenomena of neutron stars.
This possibility is intriguing, but we leave detailed analysis beyond the scope of this paper and apply the common assumption of the (locally) isotropic crust.
Within this assumption,
the elasticity of the crust matter can be described by two (effective) elastic parameters: shear and bulk moduli. Despite that, as for terrestrial materials, the assumption of isotropic crust is not enough to calculate these parameters straightforwardly. In contrast, there is a fundamental problem in such calculations: if the elastic properties of crystallites are strongly anisotropic (that is the case for the crust within the one-component approximation \cite{Fuchs36,oi90}), the effective shear and bulk moduli
depend on the actual shapes and correlations in the orientation of crystallites and therefore cannot be calculated precisely in the absence of detailed information on the polycrystalline microstructure \cite{ll_elast}.

Hence, estimates for the effective elasticity parameters are given within some assumptions.
Historically, the first approach was proposed by Voigt \cite{Voigt_1887} for terrestrial science; it was based on the assumption of uniform strain in all crystallites. The first estimate of the effective shear modulus for the neutron star crust was obtained in Ref.\ \cite{oi90} via another approach (averaging of the shear wave velocity), which leads to the same result as Voigt's assumption (Eq. \ref{mu_V}). Within the Coulomb crystal approximation, the Voigt estimate for the effective shear modulus can be calculated analytically, leading to a simple explicit expression;  in particular,  the latter indicates that the Voigt estimate weakly depends on the actual microscopic structure \cite{C21_elastCoins}; the screening corrections can also be taken into account \cite{Chugunov22_elast_screen}.
The Voigt estimate for the effective shear modulus is widely used in astrophysical applications (e.g., to study torsional oscillations \cite{YakovkevDG23,KY20,Strohmayer_etal91,Passamonti_Pons_16}), and, in fact, it was generally believed to be accurate within the neutron star literature, until \cite{kp15} points to other approaches developed in terrestrial material science.  

Let us return to the review of approaches in terrestrial material science.
The second method to calculate the elastic properties of polycrystalline matter was proposed by Reuss \cite{Reuss_29}.
It assumes that the stress (not the strain, as in the Voigt approach) is constant along the polycrystalline matter. 
As shown by Hill \cite{Hill_1952},
the effective elastic moduli lie between the Reuss and Voigt estimates for any polycrystalline structure. In the following, we refer to them as the Voigt-Reuss bounds.
Another interesting result of material science, pointed and applied in neutron star applications by \cite{kp15}, is the self-consistent theory \cite{Kroner58, Eshelby61}.
This theory provides a certain estimate for the effective shear modulus.
However, 
the self-consistent estimate is, in fact, based on several additional assumptions (see, e.g., \cite{deWit_sc_model} for more details): (a) the orientation of the crystallites is uniform and random and (b) the shape of the crystallites is spherical. The assumption (a) seems reasonable (at least due to the absence of any reliable information on the correlations), but the assumption (b) can be too strong and can affect the result (see, e.g., \cite{Diff_shape_inclus} for the detailed study of the similar problem for composites).

In this paper we,
for the first time,
consider the effective elastic properties of the neutron star crust within the variational approach suggested by Hashin and Shtrikman \cite{HS62_a,HS62_b} and widely applied in terrestrial sciences (see, for example, \cite{Kube_Arguelles_16, Brown_15} and references therein).
This approach keeps the assumption (a) mentioned above -- the random orientation of the crystallites, but has no assumption on the crystallite shapes. Generally, the latter does not allow to predict certain values for the elastic moduli, rather regions (Hashin-Shtrikman bounds) of possible values of effective shear and bulk modulus are obtained.
Due to cubic symmetry of the crystallites, the bulk modulus $K$ is already determined by the Voigt-Reuss bounds (they coincide for cubic symmetry), and
the application of the Hashin-Shtrikman approach to the neutron star crust is free of numerical difficulties (mentioned, e.g., in Ref. \cite{Brown_15}), being effectively reduced to explicit algebraic formulae for the effective shear modulus (Section~\ref{Sec:HS_bounds}).
The numerical results are presented in Section \ref{Sec:Num}, which is separated for subsections, considering the following models: the static one-component Coulomb crystal, the Yukawa crystal, the
Coulomb crystal with
nuclei motions, and binary Coulomb crystals.
In all cases, composition for all grains assumed to be the same.
Section~\ref{Sec:Res_concl} summarizes our results.

\section{Elastic properties of polycrystalline matter: general formulae}
\label{Sec:HS_bounds}

As discussed in the Introduction, it is not trivial to calculate the elastic properties of polycrystalline matter, and, furthermore, the result can depend on the actual configuration of crystallites. 
The latter is generally not known (especially for the case of the neutron star crust considered here), and calculations of the elastic properties are performed within assumptions.
In this section, we briefly review these approaches and present the respective formulae.

For the sake of clarity, let us recall that the matter of the neutron star crust has finite pressure $P$ at an undeformed state, so the elastic theory for such a system is more complicated than the standard textbook
version \cite{ll_elast}. Namely, the neutron star crust is characterised by two different elasticity tensors of rank four. One, denoted as $S_{ijkl}$, describes the perturbation of the energy $E$ of matter element, which initially has volume $V$:
\begin{equation}
    \frac{\delta E}{V} = -P u_{ii}+\frac{1}{2}S_{ijkl}u_{ij}u_{kl}.
\end{equation}
The second is the stress-strain elasticity tensor $B_{ijkl}$; it describes a perturbation of the stress tensor \begin{equation}
    \delta \sigma_{ij} = \frac{1}{2}B_{ijkl} \left( u_{kl}+u_{lk}\right).
\end{equation} 
Here $u_{ij}$ is the displacement gradient of the matter element (see \cite{Wallace67} for details and \cite{Chugunov22_elast_screen} for a brief summary). In this paper, we consider the stress-strain elasticity tensor $B_{ijkl}$, which retains the same symmetry properties as in the elasticity theory for zero pressure in the undeformed state.

\subsection{Elasticity of crystallites with cubic symmetry and isotropic matter}
\label{Sec:Elast_isotr_cubic}
In this paper we consider polycrystalline matter consisting of crystallites with cubic symmetry; the elastic properties of such crystallites can be described by the stress-strain elasticity tensor which in the principal axes and the Voigt notation
has the following form:
\begin{equation}
\bm B=
\begin{pmatrix}
c_{11} & c_{12} & c_{12} & 0 & 0 & 0 \\
c_{12} & c_{11} & c_{12} & 0 & 0 & 0 \\
c_{12} & c_{12} & c_{11} & 0 & 0 & 0 \\
0 & 0 & 0 & c_{44} & 0 & 0 \\
0 & 0 & 0  & 0 & c_{44} & 0 \\
0 & 0 & 0  & 0 & 0 & c_{44}
\end{pmatrix}.
\end{equation}
The bulk modulus, which describes the response to isotropic compression,  is defined as
$K=\frac{1}{3} \left(c_{11}+2c_{12} \right)$, while $c_{44}$ and $\frac{1}{2}(c_{11}-c_{12})$ describe the response to shear deformation at constant volume.

\subsection{Voigt and Reuss bounds}
Estimates by Voigt \cite{Voigt_1887} and Reuss \cite{Reuss_29} assume an equal volume fraction, occupied by crystallites with all orientations, supplemented by artificial simplifying assumption on the deformation field inside polycrystalline matter, the latter is required to obtain a final result analytically in explicit form. 

Voigt \cite{Voigt_1887} suggested to estimate the effective elastic properties assuming uniform strain in all crystallites. 
For cubic symmetry, it leads to 
\begin{eqnarray}
    \mu_\mathrm{V}&=&\frac{1}{5}\left(c_{11}-c_{12}+3c_{44}\right),
    \label{mu_V}\\
    K_\mathrm{V}&=& \frac{1}{3} \left(c_{11}+2c_{12} \right)\label{K_V} .
\end{eqnarray}
This result can be easily derived using rotational invariants of the elasticity tensor (e.g., \cite{Blaschke17_elast,Chugunov22_elast_screen}).
As shown by Hill \cite{Hill_1952}, $\mu_\mathrm{V}$ and $K_\mathrm{V}$ presents an upper bound for the effective shear and bulk modulus of polycrystalline matter. It is physically transparent: if we abandon the assumption of uniform strain, the strain field relaxes to decrease the deformation energy, and the reduced energy would correspond to the lower effective modulus.

The estimate by Reuss \cite{Reuss_29} is based on the assumption of uniform stress.
For crystallites with cubic symmetry, it leads to
\begin{eqnarray}
    \mu_\mathrm{R}&=&\frac{5\left(c_{11}-c_{12} \right)c_{44}}{4c_{44}+3\left(c_{11}-c_{12} \right)}.
        \label{mu_R}\\
    K_\mathrm{R}&=&\frac{1}{3} \left(c_{11}+2c_{12} \right) \label{K_R} 
\end{eqnarray}
This result can also be obtained using rotational invariants of the reverse elasticity tensor (e.g., \cite{Blaschke17_elast}) and, as shown by Hill \cite{Hill_1952}, the Reuss estimate presents a lower bound for effective moduli of polycrystalline matter.
The proof of this statement is analogous to the proof that the Voigt estimate provides an upper bound for elastic parameters  \cite{Blaschke17_elast}. Instead of considering elastic energy, one should examine its Legendre transform  -- the complementary energy 
-- which depends on the stress tensor field through the compliance tensor (the inverse elasticity tensor). The complementary energy attains a minimum at the true stress field and is overestimated under the Reuss assumption of uniform stress.
Similarly, just as one estimates the effective shear modulus based on the elastic energy of polycrystalline materials, one can estimate the effective compliance tensor from the complementary energy. As previously mentioned, it leads to overestimation under the Reuss assumption. The final step of the proof is to observe that an overestimated effective compliance tensor implies that its inverse, the effective elastic tensor, is underestimated (see, e.g., \cite{HS62_a, HS62_b}).

Note that 
only 
for crystallites with cubic symmetry, the Voigt and Reuss estimates for the bulk modulus coincide ($K_\mathrm{V}=K_\mathrm{R}$). Since they represent the upper and lower bounds, the effective bulk modulus $K_\mathrm{eff}=K_\mathrm{V}=K_\mathrm{R}=K$ is well determined. This result is natural: cubic symmetry is preserved under uniform compression described 
by $K$. In this case, the assumptions underlying the Voigt and Reuss estimates lead to essentially the same deformation field in both cases (isotropic compression on microscopic scale), which adequately represents the actual deformation field of polycrystalline matter under isotropic compression. 
Thus, in the following, we concentrate on the constraints on the effective shear modulus.

Having upper and lower bounds for the effective shear modulus, Hill \cite{Hill_1952} suggested applying an arithmetic mean of these estimates
\begin{eqnarray}
    \mu_\mathrm{Hill}&=&\frac{\mu_\mathrm{V}+\mu_\mathrm{R}}{2}.
        \label{mu_H}
\end{eqnarray}

\subsection{Hashin and Shtrikman bounds}

It can be shown that the Voigt-Reuss bounds cannot be improved in a general case (see, e.g.,  \cite{Avellaneda_Milton89} for consideration of specific ordered structures, where these bounds are achieved). Stronger bounds can be obtained only within additional assumptions (knowledge) on the crystallite structure of polycrystalline matter \cite{Kroner_77}.

Here we apply an approach, suggested by Hashin and Shtrikman \cite{HS62_b} within a natural assumption that there are no correlations between crystallites (or, more formally, 
the correlation functions of any order show no signs of violation of isotropy and homogeneity \cite{Kroner_77}). The approach is based on the consideration of the difference between actual (microscopical) deformations of polycrystalline matter and `reference matter'
with given isotropic elastic properties. They formulated the variational approach, which allowed them to suggest refined upper and lower bounds for the polycrystalline elasticity coefficients; each bound corresponds to the special choice of the elastic properties of the reference matter. 
Namely, Hashin and Shtrikman demonstrate that the variational functional reaches 
a stationary point 
at the accurate displacement field of the polycrystal;
for crystallites with cubic symmetry of elastic properties, this stationary point is the maximum if the shear modulus of reference matter $\mu_0<\mathrm{min}(c_{44},(c_{11}-c_{12})/2)$, and the minimum if $\mu_0>\mathrm{max}(c_{44},(c_{11}-c_{12})/2)$.
The Hashin-Shtrikman bounds correspond to the choice $\mu_0=c_{44}$ and $\mu_0=(c_{11}-c_{12})/2$ and some assumption of the displacement field; however, the respective derivations are far from trivial \cite{HS62_b}.
Here we present just a final expression 
for isotropic and homogeneous polycrystalline matter composed of crystallites with cubic symmetry of elastic properties. The first  bound is
\begin{eqnarray}
\mu_\mathrm{HS}^{(1)}&=&\frac{1}{2}\left(c_{11}-c_{12} \right) +3\left(\frac{10}{2c_{44}-c_{11}+c_{12}} 
\right.
\label{HSl_gen}\\
&&\left.+\frac{24\left(K+c_{11}-c_{12}\right)}{5\left(c_{11}-c_{12} \right)\left(3K+2c_{11}-2c_{12} \right)} \right)^{-1}
\nonumber
\end{eqnarray}
 and the second bound is
\begin{equation}
\mu_\mathrm{HS}^{(2)}=c_{44}+\left(\frac{5}{c_{11}-c_{12}-2c_{44}}+\frac{9 \left( K+2c_{44}\right)}{5c_{44}\left( 3K+4c_{44}\right)} \right)^{-1}.
\label{HSup_gen}
\end{equation}
In the case $c_{44}>(c_{11}-c_{12})/2$, $\mu_\mathrm{HS}^{(1)}$ is a lower bound and  $\mu_\mathrm{HS}^{(2)}$ is an upper bound.
For the opposite case $c_{44}<(c_{11}-c_{12})/2$, the bounds exchange their roles, i.e.\ $\mu_\mathrm{HS}^{(1)}$ defines the upper bound and $\mu_\mathrm{HS}^{(2)}$ -- the lower bound.

Recall that the effective bulk modulus of isotropic and homogeneous polycrystalline matter composed of crystallites with cubic symmetry coincides with the bulk modulus $K$ of crystallites (see, for example,~\cite{Hill_1952, deWit_sc_model}).
In the neutron star crust, the pressure and the bulk modulus $K$ are mainly determined by the contribution of electrons (and neutrons in the inner crust, e.g., \cite{CSC25} and references therein), so it is possible to use an approximation $K \gg c_{44},c_{11}-c_{12}$. In this case, the formulae for the Hashin-Shtrikman bounds can be a bit simplified: 
\begin{eqnarray}
\mu_\mathrm{HS}^{(1)}&=&\frac{1}{2}\left(c_{11}-c_{12} \right)\nonumber\\
&+&3\left(\frac{10}{2c_{44}-c_{11}+c_{12}}+\frac{8}{5\left(c_{11}-c_{12} \right)} \right)^{-1},
\label{HSl}
\\
\mu_\mathrm{HS}^{(2)}&=&c_{44}+\left(\frac{5}{c_{11}-c_{12}-2c_{44}}+\frac{3}{5c_{44}} \right)^{-1}.
\label{HSup}
\end{eqnarray}

Note that in the form (\ref{HSl}, \ref{HSup}) the bounds do not depend on $K$, but only on the elastic coefficients, which describe the response of crystallites to shear deformations.
In the case of the neutron star crust, considered here, the latter are determined by the Coulomb interaction of nuclei between themselves and the (electron) background, being determined by the same expressions for the outer and for the inner crust (only in the innermost layers of inner crust one should take into account effects associated with finite size of nuclei, see \cite{ZC23_shear_crust,Xia_ea23_Elast}).
As a result, the Hashin-Shtrikman bounds can be described in the whole crust within a universal approach (see Section~\ref{Sec:Num}).

\subsection{Self-consistent theory}
To obtain an estimate, rather than bounds for the effective shear modulus, one needs to employ even more detailed assumptions. In particular, it can be done within the self-consistent approach, which is based on the Eshelby’s theory of elastic inclusions and inhomogeneities  \cite{Eshelby57,Eshelby61}.

The theory of the self-consistent approach is not trivial, and here we present it schematically, referring the readers to \cite{deWit_sc_model} for details.
The general concept of the self-consistent estimate is the assumption that the overall response of the polycrystalline matter is the same as the average response of the crystallites.
The implementation of this idea can be divided into several steps.
In the first step, one considers a stress applied to a large domain of elastic material with isotropic elastic properties, containing an {\it inhomogeneity} -- subdomain with anisotropic elastic properties; the inhomogeneity mimics a crystallite.
The stress field in the system is disturbed by the inhomogeneity. Eshelby suggested an {\it equivalent inclusion} approach to analyse this disturbance. It is based on a consideration of a similar problem for the auxiliary system, which is called an {\it inclusion}: a large domain of isotropic elastic material with a subdomain with the same isotropic elastic moduli, but with finite strain in a stress-free state. In the case of the ellipsoidal subdomain,
and the uniform stress-free strain, 
the stress perturbations in the inclusion are also uniform and
can be expressed via specific integrals that depend on the axes of the ellipsoid.
This solution helps to determine the stress disturbance 
for the inhomogeneity problem.
In the next step, the stress disturbance is averaged over orientations of inhomogeneity (crystallite).
Finally, the self-consistency condition is imposed: the elastic properties of the material are adjusted to vanish averaged stress disturbance; the respective shear and bulk moduli represent the self-consistent estimates.

As follows from the above, the predicted self-consistent elastic moduli depend on the assumed shape of the crystallites. The widely applied 
approach corresponds to the assumption of spherical crystallites that allows to derive 
the above mentioned integrals analytically.
If, in addition, the elastic properties of crystallites have cubic symmetry, the self-consistent approach is especially simple: it results in a cubic equation for the self-consistent shear modulus \cite{Kroner58}.
This result was applied in \cite{kp15} to estimate the effective shear modulus of the neutron star crust. Namely, making use the relations $K \gg c_{44}$ and $K \gg c_{11}-c_{12}$ authors of Ref.\ \cite{kp15} reduced the cubic equation to quadratic, solved it, and wrote the self-consistent effective shear modulus as
\begin{equation}
\mu_\mathrm{sc}=\frac{c_{44}}{6} \left(1+\sqrt{1+12\frac{c_{11}-c_{12}}{c_{44}}} \right).
\label{sc}
\end{equation}
They also applied this equation to calculate $\mu_\mathrm{sc}$ for the neutron star crust within the one-component Coulomb crystal approximation (see Subsection \ref{SubSec:OCP} below).

\section{Constraints on the shear modulus of the neutron star crust: numerical results}
\label{Sec:Num}
In this section we consider bounds for the effective shear modulus of the neutron star crust 
within several models, applied to calculate elastic properties of crystallites; the results are applicable to both the outer and inner crusts, as well as to the cores of white dwarfs. The only exception is the innermost layers of the inner crust, where the internuclear distance becomes comparable to the size of the nuclei, and the elastic properties are affected by deformation of nuclei \cite{ZC23_shear_crust,Xia_ea23_Elast}.
For generality, most of the results are presented in appropriate dimensionless units, and in Subsection \ref{SubSec:ion_motion} we illustrate our results in physical units, using pure $^{12}$C and $^{56}$Fe matter as examples.

\subsection{One-component Coulomb crystal}
\label{SubSec:OCP}

\begin{table}
\centering
\begin{tabular}{l c c c c c c c}
\hline
\hline
  &$c_{44}$ & $c_{11}-c_{12}$
  &  $\mu_\mathrm{R}$ & $\mu_\mathrm{HS}^{-}$ & $\mu_\mathrm{sc}$ & $\mu_\mathrm{HS}^{+}$ & $\mu_\mathrm{V}$
\\
 \hline
bcc 
 & 0.1828 & 0.0490
  & 0.0510 & 0.0712 & 0.0930 & 0.1028 & 0.1195
\\
fcc 
  & 0.1853 & 0.0413 
 & 0.0443 & 0.0641 & 0.0901 & 0.1016  & 0.1194
  \\
\hline
\hline
\end{tabular}
\caption{Various estimates of the
effective shear modulus [in units of $nZ^2e^2/a$] of polycrystalline matter in the static Coulomb crystal approximation.
See text for details.
}
\label{Tabl_mu}
\end{table}

The simplest model to analyse the elastic properties of the crust is the perfect static Coulomb crystal.
Within this model, the appropriate unit for the elasticity tensor of crystallites is $Z^2 e^2n/a$, where $Ze$ is the charge of the nuclei,
$n$ is the number density of the nuclei, $a=\left(4 \pi n/3 \right)^{-1/3}$ is the radius of the ion sphere.
In these units, the elastic parameters are constants which depend only on the lattice type. They were calculated by the Ewald summation and, within this model, known very accurately (see, for instance, \cite{Fuchs36, Baiko11, Kozhberov19_binary,Kozhberov22}); rounded values of $c_{44}$ and $c_{11}-c_{12}$ are quoted in Table \ref{Tabl_mu} for completeness.

It allows us to calculate all estimates of the shear modulus mentioned in Section~\ref{Sec:HS_bounds}. The results are presented in Table~\ref{Tabl_mu}.
Note that within the Coulomb crystal model,  $\mu_\mathrm{V}=-(2/15) e_C$, where $e_C$ is the Coulomb energy density of the perfect crystal \cite{C21_elastCoins}; estimates for $\mu_\mathrm{R}$, $\mu_\mathrm{sc}$, $\mu_\mathrm{V}$ 
were also known \cite{oi90, Strohmayer_etal91, kp15, Kozhberov19_binary,Kozhberov22}.

The use of the Hashin-Shtrikman bounds allows us to increase the lower bound by $\sim 40\%$ (compared to $\mu_\mathrm{R}$) and reduce the upper bound by $\sim 15\%$ (compared to $\mu_\mathrm{V}$). The result obtained from the self-consistent theory lies between the Hashin-Shtrikman bounds, as it should be.

Let us note that elastic coefficients within static Coulomb crystal model are often quoted with extreme precision (e.g., 12 digits for $\mu_\mathrm{V}$ in \cite{C21_elastCoins}) and it is formally correct result, but only within this model. In principle, all estimates of the effective shear modulus can be calculated with arbitrary precision within the static Coulomb crystal model. However, these efforts seem rather unreasonable for application to the neutron star crust and white dwarf core due to several corrections discussed in the following subsections. Therefore, in Table \ref{Tabl_mu}, we present rounded (though still somewhat overprecise) values of the elastic moduli.

\subsection{One-component Yukawa crystal}
\label{SubSec:Yuk_cr}

Although the Coulomb crystal model is widely used to describe the crystallized layers of white dwarfs and the neutron star crust, the electron screening can reduce elastic properties by $\sim 10\%$ \cite{Baiko15}.

To calculate the screening corrections to the Coulomb crystal model, two approaches are usually applied. The first is based on the Thomas-Fermi dielectric function, leading to the screened
Coulomb interaction potential (the Yukawa potential). The second is to use the dielectric function calculated within the more accurate random phase approximation
(RPA) for relativistic electrons by \cite{Jancovici62}. 
According to Refs.\ \cite{Baiko02, kp21}, the first approach overestimates the effect of screening corrections on the electrostatic energy at least for the bcc lattice, and it is typically (but not always) the same for elastic properties \cite{Baiko12}.
In both cases, only leading-order screening corrections are successively taken into account \cite{kp21}.

Within the RPA approach, the elastic properties have complicated dependence on the density and composition, with specific features, associated with Friedel oscillations (see \cite{Baiko12} for details). Since the main goal of this paper is to illustrate the Hashin-Shtrikman approach in the application to the neutron star crust, here we limit ourselves to the Yukawa potential, leaving a detailed discussion of Hashin-Shtrikman bounds within the RPA approach for future studies. We should recall that the Yukawa approach is also applicable to other systems, such as dusty plasmas and charged colloids \cite{Fortov_ea05,Donko_ea08_Review,Klumov10_Melting,
KhrapakSA16_Yuk_thermod,
Khrapak_Klumov18_2D_Yukawa_elast,
Lu_ea22_2D_Yukawa_ShearSoftening, Beckers_ea23_DustyPlasma_Perspectives23,
Zhukhovitskii_Perevoschikov24,
Lipaev_ea25_WaveDispersion}.

Within the Yukawa model the elastic properties are known for several types of lattices \cite{Kozhberov22}. 
For each lattice, they depend on one dimensionless parameter $\kappa a$. 
For the white dwarf core and the neutron star crust, where screening is provided by degenerate electrons,  $\kappa$ is the Thomas-Fermi electron wave number:
\begin{equation}
    \kappa
    \approx 2 \sqrt{\frac{e^2}{\pi \hbar v_\mathrm{F}}}\frac{p_\mathrm{F}}{\hbar}.
    \label{TF_wn}
\end{equation}
Here, we introduce common notations for physical quantities: $p_\mathrm{F} = \hbar \left(3\pi^2 n_\mathrm{e} \right)^{1/3}$ is the Fermi momentum, $n_\mathrm{e}=Zn$ is the electron number density,
$v_\mathrm{F} = \partial E_\mathrm{F} / \partial p_\mathrm{F}$ is the Fermi velocity, $E_\mathrm{F}$ is the Fermi energy. 
Note that within the Yukawa model, the appropriate unit for the elasticity tensor can be chosen to be the same as for the Coulomb crystal: $Z^2 e^2n/a$.

\begin{figure}
    \centering
    \includegraphics[width=\columnwidth]{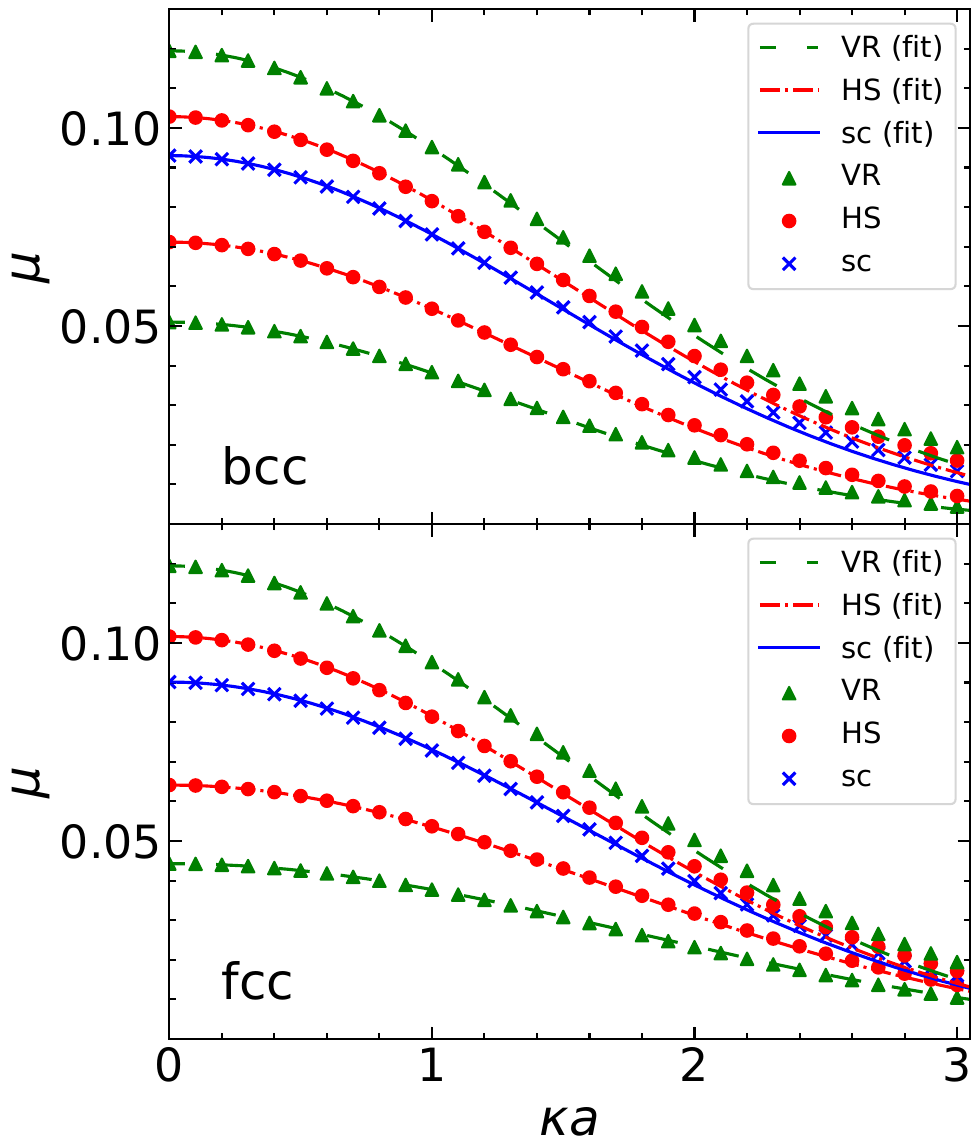}
    \caption{Effective shear modulus $\mu$  of a polycrystalline matter formed by Yukawa crystallites with bcc (the top panel) and fcc (the bottom panel) lattice as a function of the parameter $\kappa a$. $\mu$ is normalized to  $n\left(Ze \right)^2/a $. 
    Triangles (`VR') represent the Voigt (\ref{mu_V}) and Reuss (\ref{mu_R}) estimates, respectively. 
    Circles (`HS') show lower (\ref{HSl}) and upper (\ref{HSup}) Hashin-Shtrikman bounds, respectively. Crosses (`sc') indicate the self-consistent estimate (\ref{sc}). The corresponding lines reflect the fits~(\ref{Fit:mu_Yuk}).}  
    \label{Fig:mu_Yuk_bcc_fcc}
\end{figure}

The bounds for the effective shear modulus of the polycrystal, composed of Yukawa bcc or fcc crystallites, are presented in Figs.~\ref{Fig:mu_Yuk_bcc_fcc}, where we apply fits for elastic coefficients of the bcc and fcc lattice suggested by \cite{Kozhberov22} for $\kappa a \lesssim 3$
(namely, we apply Eq. (22) from the cited paper, using the parameters listed in Table II of that paper).

As expected, the Hashin-Shtrikman bounds (shown by circles) are narrower than the Voigt-Reuss bounds (shown by triangles), and the self-consistent estimate (Eq.\ \ref{sc}, crosses) lies within both bounds.
The effective shear modulus decreases with increasing $\kappa a$. This qualitative behaviour is natural because the screening corrections decrease the interaction energy of the nuclei, thereby reducing the energy perturbation associated with deformation.
The latter is equivalent to decrease of the effective shear modulus. However, for quantitative application one should keep in mind that the Yukawa potential corresponds to the leading-order perturbation theory and is not accurate enough to estimate corrections of order $(\kappa a)^4$ (e.g., \cite{kp21}).
However, this approach 
is reasonable
justified for the case of ultrarelativistic electrons, since for them the parameter
\begin{equation}
    \kappa a= \frac{18^{1/3}\alpha^{1/2}}{\pi^{1/6}} Z^{1/3}
    \approx 0.185 Z^{1/3}
\end{equation}
is rather small for any realistic $Z$ (for example, $Z=40$ leads to $(\kappa a)^2\approx 0.4$).

\begin{table}
\centering
\begin{tabular}{ l c c c c c }
\hline
\hline
 & $\mu_\mathrm{R}$ & $\mu_\mathrm{HS}^{-}$ & $\mu_\mathrm{sc}$ & $\mu_\mathrm{HS}^{+}$ & $\mu_\mathrm{V}$ \\
\hline
$d_\mathrm{Y} (bcc)$ & 0.29 & 0.27 & 0.24 & 0.23 & 0.23 \\
\hline
$d_\mathrm{Y} (fcc)$ & 0.16 & 0.18 & 0.21 & 0.22 & 0.23 \\
\hline
\hline
\end{tabular}
\caption{The fitting coefficients for different estimates of the effective shear modulus of polycrystalline matter, composed of Yukawa crystallites  with bcc and fcc lattices.}
\label{Tabl_mu_ci_Yuk}
\end{table}

To simplify application of our results, for bcc and fcc lattices we fit each estimate of the effective shear modulus of the Yukawa polycrystal by the following expression: 
\begin{equation}
    \mu \left(\kappa a\right) = \mu\left(\kappa=0\right) e^{-d_\mathrm{Y}\left(\kappa a\right)^2},
    \label{Fit:mu_Yuk}
\end{equation}
where the fitting coefficient $d_\mathrm{Y}$ is presented in Table~\ref{Tabl_mu_ci_Yuk}, the respective $\mu\left(\kappa=0\right)$ values are shown in Table~\ref{Tabl_mu}.
To provide the fit, which is the most reliable for astrophysical applications, we limit the fitting region to $\kappa a< 1$, thus for larger $\kappa a$, the fit becomes less accurate (see Fig.\ \ref{Fig:mu_Yuk_bcc_fcc}).
The practical applicability of bounds comes from two conditions:
1) The ions should be completely pressure ionized, and the electron screening must be weak. In fact, both requirements are essentially the same and can be roughly reduced to the constraint $\rho\gg 10AZ $ g cm$^{-3}$, where $A$ is the number of nucleons per nucleus \cite{hpy07}; 
2)  Eq.\ (\ref{Fit:mu_Yuk}) neglects the thermal and zero-point motion of ions, so the temperature should not be too close to the melting (see \ref{SubSec:ion_motion}  for correction for nuclei motion in the Coulomb lattice approximation, corresponding to $\kappa a=0$).

Note that the fit parameter for $\mu_\mathrm{V}$ agrees with the first-order screening correction, calculated analytically at $\kappa a\rightarrow 0$ \cite{Baiko15,Chugunov22_elast_screen}.

\subsection{One component Coulomb crystal with the
nuclei motion}
\label{SubSec:ion_motion}

The above subsections present the results within a static lattice approximation, i.e., neglecting the motion of
nuclei near the equilibrium position.

Neglecting quantum effects, the role of nucleus motions for the elastic properties was studied in the framework of Monte Carlo calculations \cite{oi90, Strohmayer_etal91} for bcc Coulomb crystals and molecular dynamics for the Yukawa crystal \cite{hh08}.
These approaches accurately apply the Coulomb (Yukawa)
interaction, but are limited by finite simulation volume, which may lead to non-negligible effects (see, e.g. Figs. 1 and 2 in \cite{hh08}).

The elastic properties of the Coulomb crystal in the thermodynamic limit (large number of nuclei) 
were analysed in Ref.\ \cite{Baiko11} within the harmonic lattice approximation. 
It allows quantum effects to be incorporated in a natural way, i.e., it is possible to take into account both thermal and zero-point oscillations of nuclei.
The quantum effects are important at $T\lesssim T_\mathrm{p}$, where  $T_\mathrm{p}=\hbar\sqrt{4 \pi n Z^2 e^2/M}/k_\mathrm{B}$ is the plasma temperature, $k_\mathrm{B}$ is the Boltzmann constant, $M$ is the mass of the nuclei
(in practice, transition to quantum asymptote for effective shear modulus occurs at $T\lesssim T_\mathrm{p}/3$, see below).

\begin{figure}
    \centering
    \includegraphics[width=\columnwidth]{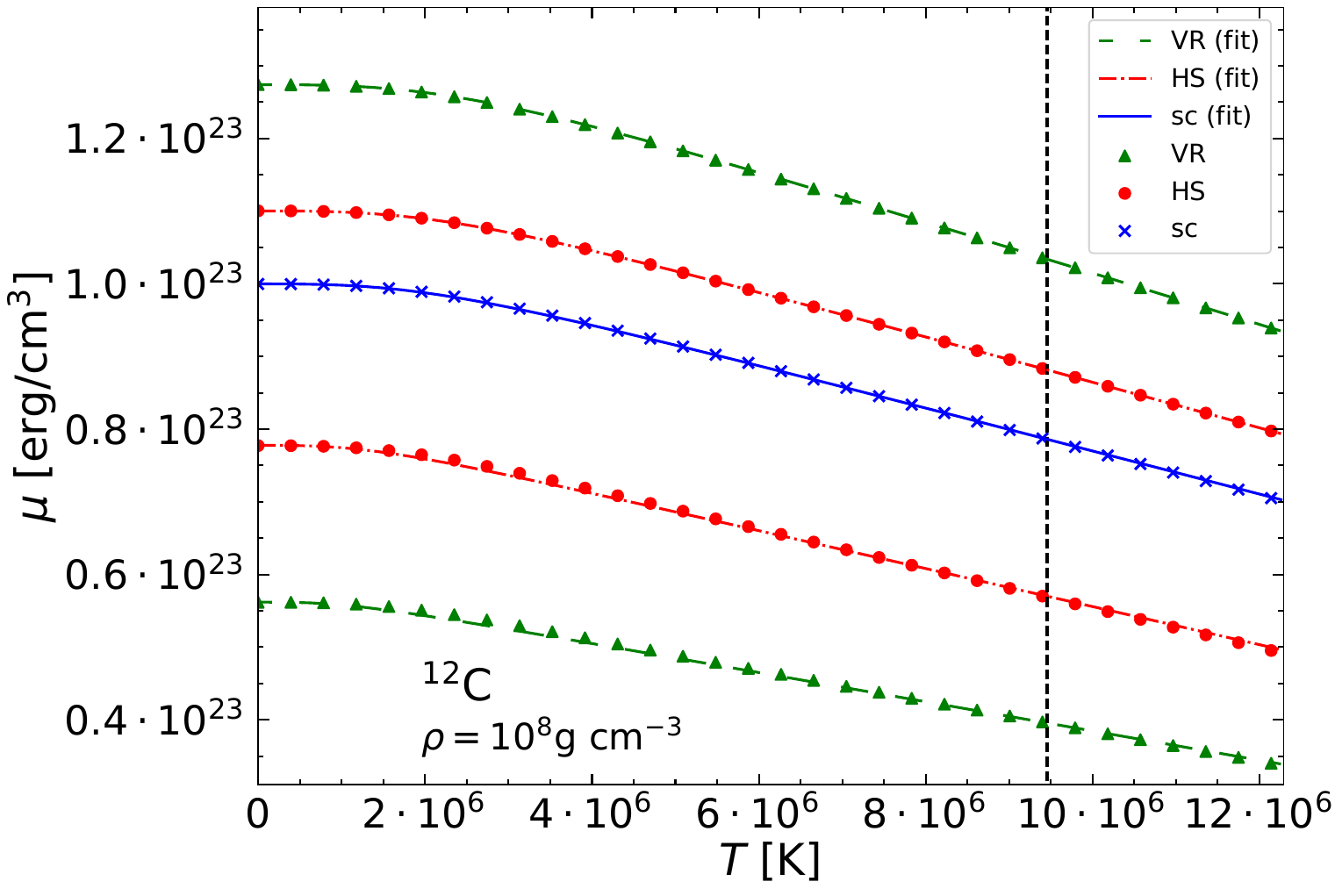}
    \caption{The effective shear modulus $\mu$ of polycrystalline
    carbon $^{12}$C at density $10^{8}$~g~cm$^{-3}$  as a function of temperature. $\mu$ was
    calculated assuming bcc crystallites and accounting for phonon corrections. 
    Triangles (`VR') represent the Voigt (\ref{mu_V}) and Reuss (\ref{mu_R}) estimates, respectively. 
    Circles (`HS') show lower (\ref{HSl}) and upper (\ref{HSup}) Hashin-Shtrikman bounds, respectively. Crosses (`sc') indicate the self-consistent estimate (\ref{sc}). The corresponding lines show Eq.\ (\ref{Fit:mu_motion}). The vertical line represents the melting temperature $T_\mathrm{m}$, calculated according to \cite{BC21}}
    \label{Fig:mu_motion_C}
\end{figure}

\begin{figure}
    \centering
    \includegraphics[width=\columnwidth]{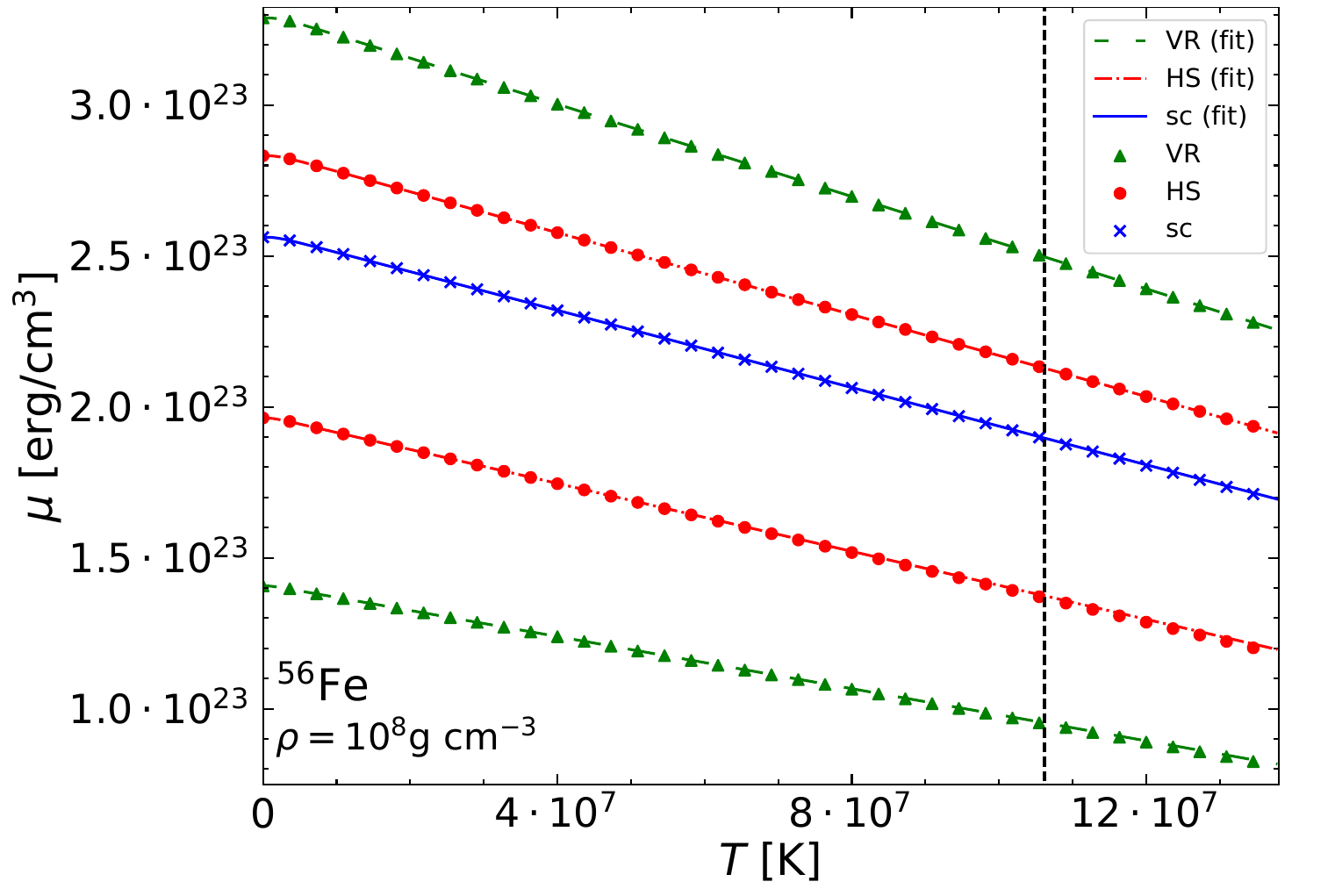}
    \caption{Similar to Fig.\ \ref{Fig:mu_motion_C}, but for  $^{56}$Fe matter. Note difference in scales with Fig.\ \ref{Fig:mu_motion_C}}
    \label{Fig:mu_motion_Fe}
\end{figure}

Ref.\ \cite{Baiko11} presents their results as 
fits for $c_{44}$ and $\mu_V$ 
via respective dimensionless parameters; the fits are applicable for any $Z$ (within the Columb crystal approximation).
To estimate $c_{11}-c_{12}$, which was not explicitly quoted in \cite{Baiko11}, we apply Eq.\ (\ref{mu_V}), i.e. we calculate it as
\begin{equation}
    c_{11}-c_{12}=5\mu_V-3 c_{44},
    \label{c_shear_estim_Baiko11}
\end{equation}
where  $c_{44}$ and $\mu_V$ are given by fits from \cite{Baiko11}.
It allows us to compute the effective shear modulus with the nucleus motion (phonon) corrections. The results are illustrated in Figs.\ \ref{Fig:mu_motion_C} and \ref{Fig:mu_motion_Fe}, plotted, respectively, for $^{12}$C and $^{56}$Fe matter; both are taken at density $10^{8}$~g~cm$^{-3}$.
Each of the plots demonstrates all estimates of the effective shear modulus as functions of temperature. The vertical dotted lines indicate the melting temperature $T_\mathrm{m}$, calculated according to \cite{BC21} (note that the melting temperature is much larger for  $^{56}$Fe). We point out that close to the melting temperature, the effective shear modulus can be affected by anharmonic corrections, which were neglected in \cite{Baiko11} and thus are not accounted for here. 

The shear modulus is predicted to be almost temperature-independent at the lowest temperatures $T\lesssim T_\mathrm{p}/3$;
this is because the nuclear motion is dominated by zero-point vibrations.
For both figures, \ref{Fig:mu_motion_C} and \ref{Fig:mu_motion_Fe}, this region corresponds to $T\lesssim 10^6$~K
(they are plotted for same density, thus $T_\mathrm{p}\propto Z/A$ is almost the same)
However, this region is clearly visible only in Fig.\ \ref{Fig:mu_motion_C}, plotted for lighter $^{12}$C nuclei.
In Fig.\ \ref{Fig:mu_motion_Fe}, plotted for heavier $^{56}$Fe nuclei, this region is almost invisible due to temperature scale:
as for Fig.\ \ref{Fig:mu_motion_C}, the temperature scale is chosen to include $T=0$~K and the melting temperature $T_\mathrm{m}$, but $T_\mathrm{m}$ is much larger for $^{56}$Fe than for $^{12}$C (see the vertical dotted lines in the plots).

\begin{table}
\centering
\begin{tabular}{l c c c c c}
\hline
\hline
  &    $\mu_\mathrm{R}$ & $\mu_\mathrm{HS}^{-}$ & $\mu_\mathrm{sc}$ & $\mu_\mathrm{HS}^{+}$ & $\mu_\mathrm{V}$

\\
 \hline
$b_\mathrm{q}$ 
 & 0.090 & 0.151
  & 0.258 & 0.305 & 0.369 
 
\\
$b_\mathrm{cl}$ 
  & 2.91 & 3.79 
 & 4.31 & 4.57 & 5.15 
  \\
\hline
\hline
\end{tabular}
\caption{The fitting coefficients for the analytical expression (\ref{Fit:mu_motion}) for nuclear motion corrections in the considered effective shear modulus estimates of polycrystalline matter composed of the Coulomb crystal.
}
\label{Tabl_mu_ph}
\end{table}

To simplify numerical applications and being based on the fitting expressions from \cite{Baiko11}, we provide an analytical approximations for estimates of the effective shear modulus with the phonon corrections:
\begin{equation}
    \mu(T) =\mu(T=0) -
    nk_\mathrm{B}T_\mathrm{p} 
    \left[
          b_\mathrm{q}^3
           +b_\mathrm{cl}^3
               \left(\frac{T}{T_\mathrm{p}}\right)^3       
    \right]^{1/3}.
    \label{Fit:mu_motion}
\end{equation}
Here the first term does not include effects of the phonon corrections.
The coefficients $b_\mathrm{q}$ and $b_\mathrm{cl}$ are quoted in Table \ref{Tabl_mu_ph}.
Their numerical values were derived by considering asymptotes of approximations from \cite{Baiko11} (supplemented with Eq.\ \ref{c_shear_estim_Baiko11}) in quantum ($T\ll T_\mathrm{p}$) and classical ($T\gg T_\mathrm{p}$) limits and adjusting $b_\mathrm{q}$ and $b_\mathrm{cl}$
to reproduce these asymptotes (note that the phonon formalism applied in \cite{Baiko11} allows to calculate the first order corrections, but neglects high order anharmonic contributions, thus an accurate treatment of next-order corrections to the effective shear modulus would be an excess of accuracy). 
Namely, $b_\mathrm{q}$ is adjusted to reproduce the correction in the quantum limit ($T\ll T_\mathrm{p}$), while $b_\mathrm{cl}$ is adjusted for the classical limit ($T\gg T_\mathrm{p}$).
By construction,
our approximation for the Voigt estimate coincides with formulae suggested by \cite{Baiko11}; the respective $b_{\mathrm{q},\mathrm{cl}}$ are quoted in Table  \ref{Tabl_mu_ph} for completeness.  
The estimates provided by (\ref{Fit:mu_motion}) are represented by lines in Figs.\ \ref{Fig:mu_motion_C}, \ref{Fig:mu_motion_Fe}.

It is worth noting that Eq.~(\ref{Fit:mu_motion}) is derived assuming fully ionised nuclei and neglecting the electron screening, i.e.\ $\rho\gg 10AZ $ g cm$^{-3}$ (see~\ref{SubSec:Yuk_cr}), it is also based on the harmonic approximation, so it can be not very accurate at temperatures close to the melting temperature, where anharmonic corrections can be important \cite{Baiko11, BC21}. 

To estimate effective shear moduli, accounting for simultaneously screening and nuclei motion, one can apply Eq.\ (\ref{Fit:mu_motion}) with $\mu(T=0)$ given by Eq.\ (\ref{Fit:mu_Yuk}). This approach allows us to take into account first-order corrections for screening and nucleus motion, neglecting the cross-effect (which is of second order).

\subsection{Binary Coulomb crystal}
\label{SubSec:binary_crystal}

The previous subsections consider the effective shear modulus within the widely used one-component approximation. In the case of the crust of the isolated neutron star, it relies on the cold-catalysed hypothesis, which assumes that the crust composition evolves to the ground state during initial stages of the neutron star life. In this case, the composition of nuclei is predicted to be one-component in the vast majority of the crust layers (e.g., \cite{cf16_mix}). However, it is generally believed that the crystallized layers of white dwarfs and the accreted neutron star crust consist of several types of nuclei in each layer and multicomponent crystals can be formed.
Currently, studies of the elastic properties of Coulomb crystals beyond the one-component approximation are limited to binary (two-component) ones.
Binary lattices can be characterized by the charge ratio of the components $R_Z=Z_2/Z_1$ (for definiteness, the charge of the nuclei of the second type $Z_2$ assumed to be larger than charge of the nuclei of the first type $Z_1$), and the number fraction of the nuclei of the second type $x_2$.

Ref.\ \cite{IH_binary_cryst} analysed
the disordered Coulomb lattice of two types: with substitutional disorder (for $R_Z\le 3$) and interstitial disorder (for $R_Z\ge3$).
The elastic properties for lattices with substitutional disorder are predicted to be rather isotropic (except for the case of the binary charge ratio $R_Z=4/3$, and $x_2=0.1$ for $R_Z=2$; see table 2 in \cite{IH_binary_cryst} for details).
Lattices with interstitial disorder were also predicted to be almost isotropic for $x_2 R_Z\lesssim 2.5$ (see table 3 in \cite{IH_binary_cryst} for details).
Eqs.\ (\ref{HSl}-\ref{HSup}) allows to compute the Hashin-Shtrikman bounds for random lattices from data in tables 2 and 3 of \cite{IH_binary_cryst}  and we do not quote these results here for brevity.

Ref.\ \cite{Kozhberov19_binary} analyse elastic properties of the ordered binary body-centred cubic lattice,
similar to the CsCl lattice:
nuclei of each type are arranged into the cubic lattice, forming thus two sublattices, shifted in such a way to locate nuclei of type '2' in a centre of a cube, formed by the nuclei of type '1' (obviously, at the same time, nuclei of type '1' are located in the centres of cubes, formed by nuclei of type '2'); following \cite{Kozhberov19_binary}, we refer to this lattice as sc2. Obviously, in the sc2 lattice the number of nuclei of each type is equal ($x_2=1/2$).
For $R_Z>3.6$ the sc2 lattice becomes unstable with respect to specific phonon modes \cite{kb12} and thus the discussion of elastic properties is meaningless.

\begin{figure}
    \centering
    \includegraphics[width=\columnwidth]{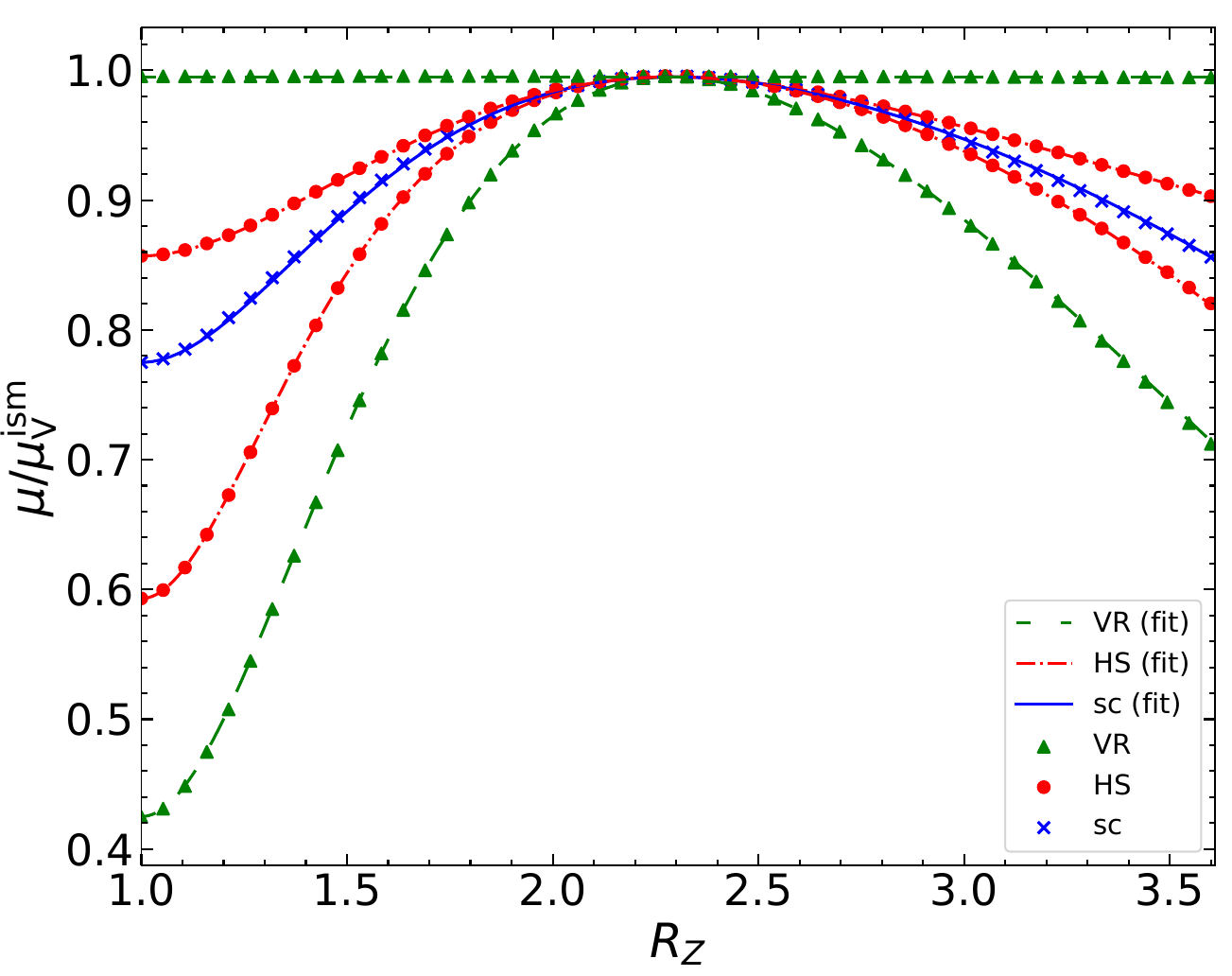}
    \caption{Effective shear modulus sc2 of the crystal $\mu$ in units of the Voigt estimate (\ref{mu_V}), as a function of the parameter $R_Z$.
    Triangles (`R') represent the Reuss estimate (\ref{mu_R}) , respectively. 
    Circles (`HS') show lower (\ref{HSl}) and upper (\ref{HSup}) Hashin-Shtrikman bounds, respectively. Crosses (`sc') indicate the self-consistent estimate (\ref{sc}). The corresponding lines reflect the fits~(\ref{Fit:mu_binary}).}
    \label{Fig:mu_binary}
\end{figure}

The estimates of the effective shear modulus for the sc2 lattice as a function of $R_Z$ are presented in Fig.~\ref{Fig:mu_binary}. The normalization is chosen to be the Voigt averaged shear modulus, derived within the ion sphere model \cite{Salpeter54} using the linear mixture rule \cite{HV76,oih93}:
\begin{equation}
   \mu_\mathrm{V}^\mathrm{ism}=\frac{3 n}{50} 
   \frac{\left(Z_1^{5/3}+Z_2^{5/3}\right) e^2}{a_e},
   \label{mu_v_ism}
\end{equation}
where $n$ is the total number density of nuclei and $a_e\equiv\left[2\pi (Z_1+Z_2) n/3]\right]^{-1/3}$ is the electron sphere radius. Note that this estimate can be done analytically  \cite{Kozhberov19_binary,C21_elastCoins}.

As noted in \cite{Kozhberov19_binary}, for $R_{Z}=R_{Z,\ \mathrm{sym}} \approx 2.29$ the elastic properties of the sc2 lattice become isotropic [i.e., $c_{44} = (c_{11}-c_{12})/2$]. As a result, all estimates of the effective shear modulus of polycrystalline matter lead to the same result $\mu_\mathrm{eff}=c_{44} = (c_{11}-c_{12})/2=\mu_\mathrm{V}$ and all estimates
in Fig.~\ref{Fig:mu_binary} coincide.

\begin{table}
\centering
\begin{tabular}{ l c c c c }
\hline
\hline
 & $\mu_\mathrm{R}$ & $\mu_\mathrm{HS}^{(1)}$ & $\mu_\mathrm{sc}$ & $\mu_\mathrm{HS}^{(2)}$  \\
\hline
$b_0$ & 0.425 & 0.593 & 0.775 & 0.857  \\
\hline
$b_1$ & 0.36 & 0.83 & 0.38 & 0.11  \\
\hline
$b_2$ & $1 \cdot 10^{-3}$ & $7 \cdot 10^{-3}$ & $-1 \cdot 10^{-2}$ & $-2 \cdot 10^{-3}$ \\
\hline\hline
\end{tabular}
\caption{The fitting coefficients for different estimates of the effective shear modulus of the binary Coulomb crystal.}
\label{Tabl_mu_ci_bin}
\end{table}

To simplify application of the results, we fit each estimate of the effective shear modulus of the binary Coulomb crystal.
The Voigt estimate is well described by $\mu_\mathrm{V}=b_0^\mathrm{V} \mu_\mathrm{V}^\mathrm{ism}$, where $b_0^\mathrm{V}\approx0.995$. 
Other estimates were fitted as:
\begin{equation}
  \frac{\mu}{\mu_\mathrm{V}^\mathrm{ism}}=
   b_0^\mathrm{V}-\frac{(b_0^\mathrm{V}-b_0)}{1+b_1\left(r_Z^-\right)^2+b_2\left(r_Z^-\right)^4}
   \left(\frac{r_Z^+}
   {r_{Z,\ \mathrm{sym}}^{+}}-1\right)^2.
    \label{Fit:mu_binary}
\end{equation}
Here, to ensure that fitting expressions reproduce obvious symmetry of the elastic coefficients of sc2 lattice with respect to the transformation $R_Z \leftrightarrow 1/R_Z$,  we introduced: 
\begin{eqnarray}
    r_Z^{-}&=&R_Z-R_Z^{-1},\\
    r_Z^{+}&=&R_Z+R_Z^{-1}-2,\\
    r_{Z,\ \mathrm{sym}}^{+}&=&R_{Z,\ \mathrm{sym}}+R^{-1}_{Z,\ \mathrm{sym}}-2.
\end{eqnarray}
The parameter
$b_0$ is the ratio of the respective shear modulus estimate to $\mu_\mathrm{V}^\mathrm{ism}$ for the one-component Coulomb crystal.
It can be easily derived from Table \ref{Tabl_mu}, but, for convenience, it is quoted in the Table~\ref{Tabl_mu_ci_bin}  along with the fitting coefficients $b_{1}$ and $b_{2}$.
Note that the fitting expression (\ref{Fit:mu_binary}) for HS bounds gives values $\mu_\mathrm{HS}^{(1)}$ and $\mu_\mathrm{HS}^{(2)}$, corresponding to Eqs.\ (\ref{HSl}) and (\ref{HSup}), respectively, for all $R_Z$.
For $R_Z<R_{Z,\ \mathrm{sym}}$, $\mu_\mathrm{HS}^{(1)}$ is the lower bound, while $\mu_\mathrm{HS}^{(2)}$ is the upper bound.
For $R_Z>R_{Z,\ \mathrm{sym}}$ they exchange their roles, i.e., $\mu_\mathrm{HS}^{(1)}$ becomes the upper bound, while $\mu_\mathrm{HS}^{(2)}$ becomes the lower bound.
For $R_Z=R_{Z,\ \mathrm{sym}}$ all estimates of the effective shear modulus, given by the fitting expression (\ref{Fit:mu_binary}), coincide, as it should be.
We also checked that $\mu_\mathrm{sc}$ lies between $\mu_\mathrm{HS}^{(1)}$ and $\mu_\mathrm{HS}^{(2)}$ in the whole region, where sc2 is stable ($1/3.6<R_Z<3.6$ \cite{kb12}).

At the end of this subsection we would like to emphasise once again that the presented fit~(\ref{Fit:mu_binary}) is applicable only for the case of the stable sc2
lattice; 
the corrections associated with nuclei motion (thermal and zero-point vibrations) as well as screening corrections are neglected because their effect on the elastic properties of perfect two-component lattice
is not yet analysed in literature. 

\section{Summary and conclusions}
\label{Sec:Res_concl}

We straighten constraints on the effective shear modulus of polycrystalline matter of the neutron star crust and the white dwarf core by calculating the Hashin-Shtrikman bounds, making use of extensive studies of elastic properties of the stellar matter in monocrystalline approximation. Namely, within the one-component approximation, we employ well-known results for the elasticity of the perfect Coulomb crystal \cite{Fuchs36,Baiko11,Kozhberov22} and corrections to this result associated with screening \cite{Kozhberov22} and motions of nuclei \cite{Baiko11}; we also apply the elastic properties of two-component crystals analysed in Ref.\ \cite{Kozhberov19_binary}.

For the one-component Coulomb crystal, the effective shear modulus is constrained as $\mu_\mathrm{HS}^{(1)} \le \mu \le  \mu_\mathrm{HS}^{(2)}$, that is,
$\sim 50\%$ narrower than the interval imposed by the well-known Voigt-Reuss bounds $\mu_\mathrm{R}\le \mu\le\mu_\mathrm{V}$ (Section \ref{SubSec:OCP}).
The screening and nuclei motion corrections 
reduce both the upper and lower bounds for the effective shear modulus; the numerical formulae, applicable for astrophysical applications, are written in sections \ref{SubSec:Yuk_cr} and \ref{SubSec:ion_motion}. 
The constraints for the effective shear modulus of polycrystalline stellar matter with ordered two-component crystallites are analysed in Section \ref{SubSec:binary_crystal}; the numerical formulae are also written there. 
Note that with increase of the charge ratio from $R_Z=1$, the bounds become narrower and merge at $R_Z \approx 2.29$ (as mentioned in \cite{Kozhberov19_binary} for $\mu_\mathrm{sc}$ and $\mu_\mathrm{V}$), 
suggesting that at this charge ratio the effective shear modulus is well determined and independent of the microstructure of polycrystalline matter (the shape of crystallites and correlations of their orientations).
For $R_Z\ge 2.29$, the effective shear modulus bounds becomes broader with increase $R_Z$.

Let us remind that the physical reason, which allows us to constrain the effective shear modulus tighter than the Voigt-Reuss bounds, is the assumption underlying the Hashin-Shtrikman approach: absence of correlations between crystallites (or more formally, 
the correlation functions of any order show no signs of violation of isotropy and homogeneity \cite{Kroner_77}).
In fact, a formalism for obtaining even narrower bounds was developed \cite{Dederichs_Zeller_73, Kroner_77}. However, these bounds depend on the assumed shape of the crystallite \cite{Dederichs_Zeller_73}, although, as the authors claim, this dependence is weak if the crystallite is spheroid-like. Under the assumption of the spherical crystallite shape, the bounds was shown to converge to the effective shear modulus \cite{Gairola_Kroner_81, Kube_Jong_16, Kube_Arguelles_16}, predicted by the self-consistent approach for spherical crystallites \cite{Kroner58,Eshelby61}.
The latter approach was applied for astrophysical applications by \cite{kp15}.
The Hashin-Shtrikman  bounds, presented here, have no assumptions on the shape of the crystallites, being in this respect general.

In conclusion, it is worth indicating the caveats associated with the Hashin-Shtrikman approach.
The underlying assumption of uncorrelated crystallites seems to be natural, but, strictly speaking, it is not proved for stellar matter. In particular, if epitaxial growth of crystallites, suggested recently in Ref.\ \cite{Baiko24_a}, is indeed relevant for the white dwarf core and the neutron star crust, crystallites would likely grow correlated (note that the epitaxial growth can lead to the growth of crystallites in a 'deformed' state, affecting the elastic and breaking properties even for crystallites as it is; see detailed study  in Ref.\ \cite{Baiko24_b}). In this case, the effective shear modulus can, in principle, lie beyond the Hashin-Shtrikman bounds or, moreover, the elastic properties of the crust can be anisotropic at macroscopic scales (see \cite{mh24_a, mh25} for a discussion of possible astrophysical consequences).
Indeed, the example of violation of the Hashin-Shtrikman bounds for the effective bulk modulus was considered in \cite{Avellaneda_Milton89}, where the authors presented multi-layered laminate structures consisting of crystallites (layers) with non-cubic symmetry.
The elastic properties of these structures was shown to be isotropic, but the effective bulk modulus for one of the structure considered was equal to the Voigt bound, whereas for another the Reuss bound was achieved (see also references in \cite{Avellaneda_Milton89, Berryman_05, Brown_15} for a more advanced introduction to the application of such structures).
We are not aware of similar examples for the effective shear modulus, but it would not be too strange if the Voigt-Reuss bounds could be reached for some specific correlated structures, so these bounds can still be applied as the most conservative estimates of the shear modulus.
Returning to the bulk modulus, we should recall that due to cubic symmetry of crystallites and the dominance of electron and neutron (in the inner neutron star crust) contributions, the bulk modulus for stellar polycrystalline matter is well specified (see Section \ref{Sec:Elast_isotr_cubic}).

\begin{acknowledgments}

We would like to thank the anonymous referee for their thoughtful review and constructive comments, which have helped us improve the paper.
The study was supported by the baseline project FFUG-2024-0002 of the Ioffe Institute.

\end{acknowledgments}


\begin{thebibliography}{75}%
	\makeatletter
	\providecommand \@ifxundefined [1]{%
		\@ifx{#1\undefined}
	}%
	\providecommand \@ifnum [1]{%
		\ifnum #1\expandafter \@firstoftwo
		\else \expandafter \@secondoftwo
		\fi
	}%
	\providecommand \@ifx [1]{%
		\ifx #1\expandafter \@firstoftwo
		\else \expandafter \@secondoftwo
		\fi
	}%
	\providecommand \natexlab [1]{#1}%
	\providecommand \enquote  [1]{``#1''}%
	\providecommand \bibnamefont  [1]{#1}%
	\providecommand \bibfnamefont [1]{#1}%
	\providecommand \citenamefont [1]{#1}%
	\providecommand \href@noop [0]{\@secondoftwo}%
	\providecommand \href [0]{\begingroup \@sanitize@url \@href}%
	\providecommand \@href[1]{\@@startlink{#1}\@@href}%
	\providecommand \@@href[1]{\endgroup#1\@@endlink}%
	\providecommand \@sanitize@url [0]{\catcode `\\12\catcode `\$12\catcode
		`\&12\catcode `\#12\catcode `\^12\catcode `\_12\catcode `\%12\relax}%
	\providecommand \@@startlink[1]{}%
	\providecommand \@@endlink[0]{}%
	\providecommand \url  [0]{\begingroup\@sanitize@url \@url }%
	\providecommand \@url [1]{\endgroup\@href {#1}{\urlprefix }}%
	\providecommand \urlprefix  [0]{URL }%
	\providecommand \Eprint [0]{\href }%
	\providecommand \doibase [0]{https://doi.org/}%
	\providecommand \selectlanguage [0]{\@gobble}%
	\providecommand \bibinfo  [0]{\@secondoftwo}%
	\providecommand \bibfield  [0]{\@secondoftwo}%
	\providecommand \translation [1]{[#1]}%
	\providecommand \BibitemOpen [0]{}%
	\providecommand \bibitemStop [0]{}%
	\providecommand \bibitemNoStop [0]{.\EOS\space}%
	\providecommand \EOS [0]{\spacefactor3000\relax}%
	\providecommand \BibitemShut  [1]{\csname bibitem#1\endcsname}%
	\let\auto@bib@innerbib\@empty
	\bibitem [{\citenamefont {Haensel}\ \emph {et~al.}(2007)\citenamefont
		{Haensel}, \citenamefont {Potekhin},\ and\ \citenamefont {Yakovlev}}]{hpy07}%
	\BibitemOpen
	\bibfield  {author} {\bibinfo {author} {\bibfnamefont {P.}~\bibnamefont
			{Haensel}}, \bibinfo {author} {\bibfnamefont {A.}~\bibnamefont {Potekhin}},\
		and\ \bibinfo {author} {\bibfnamefont {D.}~\bibnamefont {Yakovlev}},\
	}\href@noop {} {\emph {\bibinfo {title} {Neutron Stars 1: Equation of State
				and Structure}}},\ Astrophysics and Space Science Library\ (\bibinfo
	{publisher} {Springer-Verlag},\ \bibinfo {address} {Berlin},\ \bibinfo {year}
	{2007})\BibitemShut {NoStop}%
	\bibitem [{\citenamefont {{Chamel}}\ and\ \citenamefont
		{{Haensel}}(2008)}]{ch08}%
	\BibitemOpen
	\bibfield  {author} {\bibinfo {author} {\bibfnamefont {N.}~\bibnamefont
			{{Chamel}}}\ and\ \bibinfo {author} {\bibfnamefont {P.}~\bibnamefont
			{{Haensel}}},\ }\bibfield  {title} {\bibinfo {title} {{Physics of Neutron
				Star Crusts}},\ }\href@noop {} {\bibfield  {journal} {\bibinfo  {journal}
			{Liv. Rev. Relativ.}\ }\textbf {\bibinfo {volume} {11}},\ \bibinfo {pages}
		{10} (\bibinfo {year} {2008})}\BibitemShut {NoStop}%
	\bibitem [{\citenamefont {{Ogata}}\ \emph {et~al.}(1993)\citenamefont
		{{Ogata}}, \citenamefont {{Ichimaru}},\ and\ \citenamefont {{van
				Horn}}}]{oih93}%
	\BibitemOpen
	\bibfield  {author} {\bibinfo {author} {\bibfnamefont {S.}~\bibnamefont
			{{Ogata}}}, \bibinfo {author} {\bibfnamefont {S.}~\bibnamefont
			{{Ichimaru}}},\ and\ \bibinfo {author} {\bibfnamefont {H.~M.}\ \bibnamefont
			{{van Horn}}},\ }\bibfield  {title} {\bibinfo {title} {{Thermonuclear
				Reaction Rates for Dense Binary-Ionic Mixtures}},\ }\href
	{https://doi.org/10.1086/173308} {\bibfield  {journal} {\bibinfo  {journal}
			{\apj}\ }\textbf {\bibinfo {volume} {417}},\ \bibinfo {pages} {265} (\bibinfo
		{year} {1993})}\BibitemShut {NoStop}%
	\bibitem [{\citenamefont {{Jones}}\ and\ \citenamefont
		{{Ceperley}}(1996)}]{jc96}%
	\BibitemOpen
	\bibfield  {author} {\bibinfo {author} {\bibfnamefont {M.~D.}\ \bibnamefont
			{{Jones}}}\ and\ \bibinfo {author} {\bibfnamefont {D.~M.}\ \bibnamefont
			{{Ceperley}}},\ }\bibfield  {title} {\bibinfo {title} {{Crystallization of
				the One-Component Plasma at Finite Temperature}},\ }\href
	{https://doi.org/10.1103/PhysRevLett.76.4572} {\bibfield  {journal} {\bibinfo
			{journal} {\prl}\ }\textbf {\bibinfo {volume} {76}},\ \bibinfo {pages}
		{4572} (\bibinfo {year} {1996})}\BibitemShut {NoStop}%
	\bibitem [{\citenamefont {{Potekhin}}\ and\ \citenamefont
		{{Chabrier}}(2000)}]{pc00_EOS_Solid}%
	\BibitemOpen
	\bibfield  {author} {\bibinfo {author} {\bibfnamefont {A.~Y.}\ \bibnamefont
			{{Potekhin}}}\ and\ \bibinfo {author} {\bibfnamefont {G.}~\bibnamefont
			{{Chabrier}}},\ }\bibfield  {title} {\bibinfo {title} {{Equation of state of
				fully ionized electron-ion plasmas. II. Extension to relativistic densities
				and to the solid phase}},\ }\href {https://doi.org/10.1103/PhysRevE.62.8554}
	{\bibfield  {journal} {\bibinfo  {journal} {\pre}\ }\textbf {\bibinfo
			{volume} {62}},\ \bibinfo {pages} {8554} (\bibinfo {year} {2000})},\ \Eprint
	{https://arxiv.org/abs/astro-ph/0009261} {arXiv:astro-ph/0009261 [astro-ph]}
	\BibitemShut {NoStop}%
	\bibitem [{\citenamefont {{Medin}}\ and\ \citenamefont
		{{Cumming}}(2010)}]{mc10_crystallization}%
	\BibitemOpen
	\bibfield  {author} {\bibinfo {author} {\bibfnamefont {Z.}~\bibnamefont
			{{Medin}}}\ and\ \bibinfo {author} {\bibfnamefont {A.}~\bibnamefont
			{{Cumming}}},\ }\bibfield  {title} {\bibinfo {title} {{Crystallization of
				classical multicomponent plasmas}},\ }\href
	{https://doi.org/10.1103/PhysRevE.81.036107} {\bibfield  {journal} {\bibinfo
			{journal} {\pre}\ }\textbf {\bibinfo {volume} {81}},\ \bibinfo {eid} {036107}
		(\bibinfo {year} {2010})},\ \Eprint {https://arxiv.org/abs/1002.3327}
	{arXiv:1002.3327 [astro-ph.SR]} \BibitemShut {NoStop}%
	\bibitem [{\citenamefont {Caplan}\ and\ \citenamefont {Horowitz}(2017)}]{ch17}%
	\BibitemOpen
	\bibfield  {author} {\bibinfo {author} {\bibfnamefont {M.~E.}\ \bibnamefont
			{Caplan}}\ and\ \bibinfo {author} {\bibfnamefont {C.~J.}\ \bibnamefont
			{Horowitz}},\ }\bibfield  {title} {\bibinfo {title} {Colloquium:
			Astromaterial science and nuclear pasta},\ }\href
	{https://doi.org/10.1103/RevModPhys.89.041002} {\bibfield  {journal}
		{\bibinfo  {journal} {Rev. Mod. Phys.}\ }\textbf {\bibinfo {volume} {89}},\
		\bibinfo {pages} {041002} (\bibinfo {year} {2017})}\BibitemShut {NoStop}%
	\bibitem [{\citenamefont {{Caplan}}\ \emph {et~al.}(2018)\citenamefont
		{{Caplan}}, \citenamefont {{Cumming}}, \citenamefont {{Berry}}, \citenamefont
		{{Horowitz}},\ and\ \citenamefont {{Mckinven}}}]{Caplan_etal18}%
	\BibitemOpen
	\bibfield  {author} {\bibinfo {author} {\bibfnamefont {M.~E.}\ \bibnamefont
			{{Caplan}}}, \bibinfo {author} {\bibfnamefont {A.}~\bibnamefont {{Cumming}}},
		\bibinfo {author} {\bibfnamefont {D.~K.}\ \bibnamefont {{Berry}}}, \bibinfo
		{author} {\bibfnamefont {C.~J.}\ \bibnamefont {{Horowitz}}},\ and\ \bibinfo
		{author} {\bibfnamefont {R.}~\bibnamefont {{Mckinven}}},\ }\bibfield  {title}
	{\bibinfo {title} {{Polycrystalline Crusts in Accreting Neutron Stars}},\
	}\href {https://doi.org/10.3847/1538-4357/aac2d2} {\bibfield  {journal}
		{\bibinfo  {journal} {\apj}\ }\textbf {\bibinfo {volume} {860}},\ \bibinfo
		{eid} {148} (\bibinfo {year} {2018})},\ \Eprint
	{https://arxiv.org/abs/1804.06942} {arXiv:1804.06942 [astro-ph.HE]}
	\BibitemShut {NoStop}%
	\bibitem [{\citenamefont {{Fantina}}\ \emph {et~al.}(2020)\citenamefont
		{{Fantina}}, \citenamefont {{De Ridder}}, \citenamefont {{Chamel}},\ and\
		\citenamefont {{Gulminelli}}}]{Fantina_ea20_outer}%
	\BibitemOpen
	\bibfield  {author} {\bibinfo {author} {\bibfnamefont {A.~F.}\ \bibnamefont
			{{Fantina}}}, \bibinfo {author} {\bibfnamefont {S.}~\bibnamefont {{De
					Ridder}}}, \bibinfo {author} {\bibfnamefont {N.}~\bibnamefont {{Chamel}}},\
		and\ \bibinfo {author} {\bibfnamefont {F.}~\bibnamefont {{Gulminelli}}},\
	}\bibfield  {title} {\bibinfo {title} {{Crystallization of the outer crust of
				a non-accreting neutron star}},\ }\href
	{https://doi.org/10.1051/0004-6361/201936359} {\bibfield  {journal} {\bibinfo
			{journal} {\aap}\ }\textbf {\bibinfo {volume} {633}},\ \bibinfo {eid} {A149}
		(\bibinfo {year} {2020})},\ \Eprint {https://arxiv.org/abs/1912.02849}
	{arXiv:1912.02849 [astro-ph.HE]} \BibitemShut {NoStop}%
	\bibitem [{\citenamefont {{Carreau}}\ \emph {et~al.}(2020)\citenamefont
		{{Carreau}}, \citenamefont {{Gulminelli}}, \citenamefont {{Chamel}},
		\citenamefont {{Fantina}},\ and\ \citenamefont
		{{Pearson}}}]{Carreau_ea20_inner}%
	\BibitemOpen
	\bibfield  {author} {\bibinfo {author} {\bibfnamefont {T.}~\bibnamefont
			{{Carreau}}}, \bibinfo {author} {\bibfnamefont {F.}~\bibnamefont
			{{Gulminelli}}}, \bibinfo {author} {\bibfnamefont {N.}~\bibnamefont
			{{Chamel}}}, \bibinfo {author} {\bibfnamefont {A.~F.}\ \bibnamefont
			{{Fantina}}},\ and\ \bibinfo {author} {\bibfnamefont {J.~M.}\ \bibnamefont
			{{Pearson}}},\ }\bibfield  {title} {\bibinfo {title} {{Crystallization of the
				inner crust of a neutron star and the influence of shell effects}},\ }\href
	{https://doi.org/10.1051/0004-6361/201937236} {\bibfield  {journal} {\bibinfo
			{journal} {\aap}\ }\textbf {\bibinfo {volume} {635}},\ \bibinfo {eid} {A84}
		(\bibinfo {year} {2020})},\ \Eprint {https://arxiv.org/abs/1912.01265}
	{arXiv:1912.01265 [astro-ph.HE]} \BibitemShut {NoStop}%
	\bibitem [{\citenamefont {{Arnold}}\ \emph {et~al.}(2025)\citenamefont
		{{Arnold}}, \citenamefont {{Daligault}}, \citenamefont {{Saumon}},
		\citenamefont {{B{\'e}dard}},\ and\ \citenamefont {{Hu}}}]{Arnold_etal_25}%
	\BibitemOpen
	\bibfield  {author} {\bibinfo {author} {\bibfnamefont {B.}~\bibnamefont
			{{Arnold}}}, \bibinfo {author} {\bibfnamefont {J.}~\bibnamefont
			{{Daligault}}}, \bibinfo {author} {\bibfnamefont {D.}~\bibnamefont
			{{Saumon}}}, \bibinfo {author} {\bibfnamefont {A.}~\bibnamefont
			{{B{\'e}dard}}},\ and\ \bibinfo {author} {\bibfnamefont {S.~X.}\ \bibnamefont
			{{Hu}}},\ }\bibfield  {title} {\bibinfo {title} {{Crystal nucleation rates in
				one-component Yukawa systems}},\ }\href
	{https://doi.org/10.1103/PhysRevE.111.025206} {\bibfield  {journal} {\bibinfo
			{journal} {\pre}\ }\textbf {\bibinfo {volume} {111}},\ \bibinfo {eid}
		{025206} (\bibinfo {year} {2025})},\ \Eprint
	{https://arxiv.org/abs/2503.05902} {arXiv:2503.05902 [physics.plasm-ph]}
	\BibitemShut {NoStop}%
	\bibitem [{\citenamefont {{Hamaguchi}}\ \emph {et~al.}(1997)\citenamefont
		{{Hamaguchi}}, \citenamefont {{Farouki}},\ and\ \citenamefont
		{{Dubin}}}]{hfd97}%
	\BibitemOpen
	\bibfield  {author} {\bibinfo {author} {\bibfnamefont {S.}~\bibnamefont
			{{Hamaguchi}}}, \bibinfo {author} {\bibfnamefont {R.~T.}\ \bibnamefont
			{{Farouki}}},\ and\ \bibinfo {author} {\bibfnamefont {D.~H.~E.}\ \bibnamefont
			{{Dubin}}},\ }\bibfield  {title} {\bibinfo {title} {{Triple point of Yukawa
				systems}},\ }\href {https://doi.org/10.1103/PhysRevE.56.4671} {\bibfield
		{journal} {\bibinfo  {journal} {\pre}\ }\textbf {\bibinfo {volume} {56}},\
		\bibinfo {pages} {4671} (\bibinfo {year} {1997})}\BibitemShut {NoStop}%
	\bibitem [{\citenamefont {{Baiko}}(2002)}]{Baiko02}%
	\BibitemOpen
	\bibfield  {author} {\bibinfo {author} {\bibfnamefont {D.~A.}\ \bibnamefont
			{{Baiko}}},\ }\bibfield  {title} {\bibinfo {title} {{Effect of the electron
				gas polarizability on the specific heat of phonons in Coulomb crystals}},\
	}\href {https://doi.org/10.1103/PhysRevE.66.056405} {\bibfield  {journal}
		{\bibinfo  {journal} {\pre}\ }\textbf {\bibinfo {volume} {66}},\ \bibinfo
		{eid} {056405} (\bibinfo {year} {2002})}\BibitemShut {NoStop}%
	\bibitem [{\citenamefont {{Kozhberov}}(2018)}]{Kozhberov18_Yuk}%
	\BibitemOpen
	\bibfield  {author} {\bibinfo {author} {\bibfnamefont {A.}~\bibnamefont
			{{Kozhberov}}},\ }\bibfield  {title} {\bibinfo {title} {{Electrostatic energy
				and phonon properties of Yukawa crystals}},\ }\href
	{https://doi.org/10.1103/PhysRevE.98.063205} {\bibfield  {journal} {\bibinfo
			{journal} {\pre}\ }\textbf {\bibinfo {volume} {98}},\ \bibinfo {eid} {063205}
		(\bibinfo {year} {2018})},\ \Eprint {https://arxiv.org/abs/1901.04427}
	{arXiv:1901.04427 [cond-mat.mtrl-sci]} \BibitemShut {NoStop}%
	\bibitem [{\citenamefont {Kozhberov}\ and\ \citenamefont
		{Potekhin}(2021)}]{kp21}%
	\BibitemOpen
	\bibfield  {author} {\bibinfo {author} {\bibfnamefont {A.~A.}\ \bibnamefont
			{Kozhberov}}\ and\ \bibinfo {author} {\bibfnamefont {A.~Y.}\ \bibnamefont
			{Potekhin}},\ }\bibfield  {title} {\bibinfo {title} {{Electrostatic energy of
				Coulomb crystals with polarized electron background}},\ }\href
	{https://doi.org/10.1103/PhysRevE.103.043205} {\bibfield  {journal} {\bibinfo
			{journal} {Phys. Rev. E}\ }\textbf {\bibinfo {volume} {103}},\ \bibinfo
		{pages} {043205} (\bibinfo {year} {2021})},\ \Eprint
	{https://arxiv.org/abs/2104.09964} {arXiv:2104.09964 [physics.plasm-ph]}
	\BibitemShut {NoStop}%
	\bibitem [{\citenamefont {Landau}\ \emph {et~al.}(1986)\citenamefont {Landau},
		\citenamefont {Lifshitz}, \citenamefont {Kosevich},\ and\ \citenamefont
		{Pitaevskii}}]{ll_elast}%
	\BibitemOpen
	\bibfield  {author} {\bibinfo {author} {\bibfnamefont {L.}~\bibnamefont
			{Landau}}, \bibinfo {author} {\bibfnamefont {E.}~\bibnamefont {Lifshitz}},
		\bibinfo {author} {\bibfnamefont {A.}~\bibnamefont {Kosevich}},\ and\
		\bibinfo {author} {\bibfnamefont {L.}~\bibnamefont {Pitaevskii}},\
	}\href@noop {} {\emph {\bibinfo {title} {Theory of Elasticity}}},\ Course of
	theoretical physics\ (\bibinfo  {publisher} {Butterworth-Heinemann},\
	\bibinfo {year} {1986})\BibitemShut {NoStop}%
	\bibitem [{\citenamefont {{Horowitz}}\ and\ \citenamefont
		{{Hughto}}(2008)}]{hh08}%
	\BibitemOpen
	\bibfield  {author} {\bibinfo {author} {\bibfnamefont {C.~J.}\ \bibnamefont
			{{Horowitz}}}\ and\ \bibinfo {author} {\bibfnamefont {J.}~\bibnamefont
			{{Hughto}}},\ }\bibfield  {title} {\bibinfo {title} {{Molecular Dynamics
				Simulation of Shear Moduli for Coulomb Crystals}},\ }\href@noop {} {\bibfield
		{journal} {\bibinfo  {journal} {arXiv e-prints}\ ,\ \bibinfo {eid}
			{arXiv:0812.2650}} (\bibinfo {year} {2008})},\ \Eprint
	{https://arxiv.org/abs/0812.2650} {arXiv:0812.2650 [astro-ph]} \BibitemShut
	{NoStop}%
	\bibitem [{\citenamefont {{Hoffman}}\ and\ \citenamefont
		{{Heyl}}(2012)}]{hh12}%
	\BibitemOpen
	\bibfield  {author} {\bibinfo {author} {\bibfnamefont {K.}~\bibnamefont
			{{Hoffman}}}\ and\ \bibinfo {author} {\bibfnamefont {J.}~\bibnamefont
			{{Heyl}}},\ }\bibfield  {title} {\bibinfo {title} {{Mechanical properties of
				non-accreting neutron star crusts}},\ }\href
	{https://doi.org/10.1111/j.1365-2966.2012.21921.x} {\bibfield  {journal}
		{\bibinfo  {journal} {\mnras}\ }\textbf {\bibinfo {volume} {426}},\ \bibinfo
		{pages} {2404} (\bibinfo {year} {2012})},\ \Eprint
	{https://arxiv.org/abs/1208.3258} {arXiv:1208.3258 [astro-ph.SR]}
	\BibitemShut {NoStop}%
	\bibitem [{\citenamefont {{Baiko}}(2012)}]{Baiko12}%
	\BibitemOpen
	\bibfield  {author} {\bibinfo {author} {\bibfnamefont {D.~A.}\ \bibnamefont
			{{Baiko}}},\ }\bibfield  {title} {\bibinfo {title} {{Shear Modulus of a
				Coulomb Crystal of Ions: Effects of Ion Motion and Electron Background
				Polarization}},\ }\href {https://doi.org/10.1002/ctpp.201100073} {\bibfield
		{journal} {\bibinfo  {journal} {Contributions to Plasma Physics}\ }\textbf
		{\bibinfo {volume} {52}},\ \bibinfo {pages} {157} (\bibinfo {year}
		{2012})}\BibitemShut {NoStop}%
	\bibitem [{\citenamefont {{Baiko}}(2015)}]{Baiko15}%
	\BibitemOpen
	\bibfield  {author} {\bibinfo {author} {\bibfnamefont {D.~A.}\ \bibnamefont
			{{Baiko}}},\ }\bibfield  {title} {\bibinfo {title} {{Screening corrections to
				the Coulomb crystal elastic moduli}},\ }\href
	{https://doi.org/10.1093/mnras/stv1166} {\bibfield  {journal} {\bibinfo
			{journal} {\mnras}\ }\textbf {\bibinfo {volume} {451}},\ \bibinfo {pages}
		{3055} (\bibinfo {year} {2015})},\ \Eprint {https://arxiv.org/abs/1603.04227}
	{arXiv:1603.04227 [astro-ph.SR]} \BibitemShut {NoStop}%
	\bibitem [{\citenamefont {Kozhberov}(2022)}]{Kozhberov22}%
	\BibitemOpen
	\bibfield  {author} {\bibinfo {author} {\bibfnamefont {A.~A.}\ \bibnamefont
			{Kozhberov}},\ }\bibfield  {title} {\bibinfo {title} {Elastic properties of
			Yukawa crystals},\ }\href {https://doi.org/10.1063/5.0083168} {\bibfield
		{journal} {\bibinfo  {journal} {Physics of Plasmas}\ }\textbf {\bibinfo
			{volume} {29}},\ \bibinfo {pages} {043701} (\bibinfo {year} {2022})},\
	\Eprint {https://arxiv.org/abs/https://doi.org/10.1063/5.0083168}
	{https://doi.org/10.1063/5.0083168} \BibitemShut {NoStop}%
	\bibitem [{\citenamefont {{Ogata}}\ and\ \citenamefont
		{{Ichimaru}}(1990)}]{oi90}%
	\BibitemOpen
	\bibfield  {author} {\bibinfo {author} {\bibfnamefont {S.}~\bibnamefont
			{{Ogata}}}\ and\ \bibinfo {author} {\bibfnamefont {S.}~\bibnamefont
			{{Ichimaru}}},\ }\bibfield  {title} {\bibinfo {title} {{First-principles
				calculations of shear moduli for Monte Carlo-simulated Coulomb solids}},\
	}\href {https://doi.org/10.1103/PhysRevA.42.4867} {\bibfield  {journal}
		{\bibinfo  {journal} {\pra}\ }\textbf {\bibinfo {volume} {42}},\ \bibinfo
		{pages} {4867} (\bibinfo {year} {1990})}\BibitemShut {NoStop}%
	\bibitem [{\citenamefont {{Strohmayer}}\ \emph {et~al.}(1991)\citenamefont
		{{Strohmayer}}, \citenamefont {{Ogata}}, \citenamefont {{Iyetomi}},
		\citenamefont {{Ichimaru}},\ and\ \citenamefont {{van
				Horn}}}]{Strohmayer_etal91}%
	\BibitemOpen
	\bibfield  {author} {\bibinfo {author} {\bibfnamefont {T.}~\bibnamefont
			{{Strohmayer}}}, \bibinfo {author} {\bibfnamefont {S.}~\bibnamefont
			{{Ogata}}}, \bibinfo {author} {\bibfnamefont {H.}~\bibnamefont {{Iyetomi}}},
		\bibinfo {author} {\bibfnamefont {S.}~\bibnamefont {{Ichimaru}}},\ and\
		\bibinfo {author} {\bibfnamefont {H.~M.}\ \bibnamefont {{van Horn}}},\
	}\bibfield  {title} {\bibinfo {title} {{The shear modulus of the neutron star
				crust and nonradial oscillations of neutron stars}},\ }\href
	{https://doi.org/10.1086/170231} {\bibfield  {journal} {\bibinfo  {journal}
			{\apj}\ }\textbf {\bibinfo {volume} {375}},\ \bibinfo {pages} {679} (\bibinfo
		{year} {1991})}\BibitemShut {NoStop}%
	\bibitem [{\citenamefont {{Baiko}}(2011)}]{Baiko11}%
	\BibitemOpen
	\bibfield  {author} {\bibinfo {author} {\bibfnamefont {D.~A.}\ \bibnamefont
			{{Baiko}}},\ }\bibfield  {title} {\bibinfo {title} {{Shear modulus of neutron
				star crust}},\ }\href {https://doi.org/10.1111/j.1365-2966.2011.18819.x}
	{\bibfield  {journal} {\bibinfo  {journal} {\mnras}\ }\textbf {\bibinfo
			{volume} {416}},\ \bibinfo {pages} {22} (\bibinfo {year} {2011})},\ \Eprint
	{https://arxiv.org/abs/1104.0173} {arXiv:1104.0173 [astro-ph.SR]}
	\BibitemShut {NoStop}%
	\bibitem [{\citenamefont {{Igarashi}}\ and\ \citenamefont
		{{Iyetomi}}(2003)}]{IH_binary_cryst}%
	\BibitemOpen
	\bibfield  {author} {\bibinfo {author} {\bibfnamefont {T.}~\bibnamefont
			{{Igarashi}}}\ and\ \bibinfo {author} {\bibfnamefont {H.}~\bibnamefont
			{{Iyetomi}}},\ }\bibfield  {title} {\bibinfo {title} {{Phase characteristics
				and elastic properties of binary Coulomb compounds}},\ }\href
	{https://doi.org/10.1088/0305-4470/36/22/348} {\bibfield  {journal} {\bibinfo
			{journal} {Journal of Physics A Mathematical General}\ }\textbf {\bibinfo
			{volume} {36}},\ \bibinfo {pages} {6197} (\bibinfo {year}
		{2003})}\BibitemShut {NoStop}%
	\bibitem [{\citenamefont {{Kozhberov}}(2019)}]{Kozhberov19_binary}%
	\BibitemOpen
	\bibfield  {author} {\bibinfo {author} {\bibfnamefont {A.~A.}\ \bibnamefont
			{{Kozhberov}}},\ }\bibfield  {title} {\bibinfo {title} {{Elastic properties
				of binary crystals in neutron stars and white dwarfs}},\ }\href
	{https://doi.org/10.1093/mnras/stz1151} {\bibfield  {journal} {\bibinfo
			{journal} {\mnras}\ }\textbf {\bibinfo {volume} {486}},\ \bibinfo {pages}
		{4473} (\bibinfo {year} {2019})},\ \Eprint {https://arxiv.org/abs/1912.11395}
	{arXiv:1912.11395 [astro-ph.HE]} \BibitemShut {NoStop}%
	\bibitem [{\citenamefont {{Baiko}}(2024{\natexlab{a}})}]{Baiko24_a}%
	\BibitemOpen
	\bibfield  {author} {\bibinfo {author} {\bibfnamefont {D.~A.}\ \bibnamefont
			{{Baiko}}},\ }\bibfield  {title} {\bibinfo {title} {{Liquid-phase epitaxy of
				neutron star crusts and white dwarf cores}},\ }\href
	{https://doi.org/10.1093/mnras/stae020} {\bibfield  {journal} {\bibinfo
			{journal} {\mnras}\ }\textbf {\bibinfo {volume} {528}},\ \bibinfo {pages}
		{408} (\bibinfo {year} {2024}{\natexlab{a}})},\ \Eprint
	{https://arxiv.org/abs/2312.17544} {arXiv:2312.17544 [astro-ph.SR]}
	\BibitemShut {NoStop}%
	\bibitem [{\citenamefont {{Baiko}}(2024{\natexlab{b}})}]{Baiko24_b}%
	\BibitemOpen
	\bibfield  {author} {\bibinfo {author} {\bibfnamefont {D.~A.}\ \bibnamefont
			{{Baiko}}},\ }\bibfield  {title} {\bibinfo {title} {{Elastic and breaking
				properties of epitaxial face-centered crystals in neutron star crusts and
				white dwarf cores}},\ }\href {https://doi.org/10.1002/ctpp.202400004}
	{\bibfield  {journal} {\bibinfo  {journal} {Contributions to Plasma Physics}\
		}\textbf {\bibinfo {volume} {64}},\ \bibinfo {eid} {e202400004} (\bibinfo
		{year} {2024}{\natexlab{b}})},\ \Eprint {https://arxiv.org/abs/2403.11991}
	{arXiv:2403.11991 [astro-ph.SR]} \BibitemShut {NoStop}%
	\bibitem [{\citenamefont {{Morales}}\ and\ \citenamefont
		{{Horowitz}}(2024)}]{mh24_a}%
	\BibitemOpen
	\bibfield  {author} {\bibinfo {author} {\bibfnamefont {J.~A.}\ \bibnamefont
			{{Morales}}}\ and\ \bibinfo {author} {\bibfnamefont {C.~J.}\ \bibnamefont
			{{Horowitz}}},\ }\bibfield  {title} {\bibinfo {title} {{Anisotropic neutron
				star crust, solar system mountains, and gravitational waves}},\ }\href
	{https://doi.org/10.1103/PhysRevD.110.044016} {\bibfield  {journal} {\bibinfo
			{journal} {\prd}\ }\textbf {\bibinfo {volume} {110}},\ \bibinfo {eid}
		{044016} (\bibinfo {year} {2024})},\ \Eprint
	{https://arxiv.org/abs/2309.04855} {arXiv:2309.04855 [astro-ph.HE]}
	\BibitemShut {NoStop}%
	\bibitem [{\citenamefont {{Morales}}\ and\ \citenamefont
		{{Horowitz}}(2025)}]{mh25}%
	\BibitemOpen
	\bibfield  {author} {\bibinfo {author} {\bibfnamefont {J.~A.}\ \bibnamefont
			{{Morales}}}\ and\ \bibinfo {author} {\bibfnamefont {C.~J.}\ \bibnamefont
			{{Horowitz}}},\ }\bibfield  {title} {\bibinfo {title} {{Limiting Rotation
				Rate of Neutron Stars from Crust Breaking and Gravitational Waves}},\ }\href
	{https://doi.org/10.3847/2041-8213/ad9ea7} {\bibfield  {journal} {\bibinfo
			{journal} {\apjl}\ }\textbf {\bibinfo {volume} {978}},\ \bibinfo {eid} {L8}
		(\bibinfo {year} {2025})},\ \Eprint {https://arxiv.org/abs/2410.19111}
	{arXiv:2410.19111 [astro-ph.HE]} \BibitemShut {NoStop}%
	\bibitem [{\citenamefont {Fuchs}(1936)}]{Fuchs36}%
	\BibitemOpen
	\bibfield  {author} {\bibinfo {author} {\bibfnamefont {K.}~\bibnamefont
			{Fuchs}},\ }\bibfield  {title} {\bibinfo {title} {A quantum mechanical
			calculation of the elastic constants of monovalent metals},\ }\href@noop {}
	{\bibfield  {journal} {\bibinfo  {journal} {Proc.\ of the Royal Society of
				London A: Mathematical, Physical and Engineering Sciences}\ }\textbf
		{\bibinfo {volume} {153}},\ \bibinfo {pages} {622} (\bibinfo {year}
		{1936})}\BibitemShut {NoStop}%
	\bibitem [{\citenamefont {{Voigt}}(1887)}]{Voigt_1887}%
	\BibitemOpen
	\bibfield  {author} {\bibinfo {author} {\bibfnamefont {W.}~\bibnamefont
			{{Voigt}}},\ }\bibfield  {title} {\bibinfo {title} {{Theoretische Studien
				uber die Elastizit{\"a}tsverhh{\"a}ltnisse der Kristalle.}},\ }\href@noop {}
	{\bibfield  {journal} {\bibinfo  {journal} {Abh. Kgl. Ges. Wis.
				G{\"o}ttingen}\ }\textbf {\bibinfo {volume} {34}},\ \bibinfo {pages} {1}
		(\bibinfo {year} {1887})}\BibitemShut {NoStop}%
	\bibitem [{\citenamefont {{Chugunov}}(2021)}]{C21_elastCoins}%
	\BibitemOpen
	\bibfield  {author} {\bibinfo {author} {\bibfnamefont {A.~I.}\ \bibnamefont
			{{Chugunov}}},\ }\bibfield  {title} {\bibinfo {title} {{Neutron star crust in
				Voigt approximation: general symmetry of the stress-strain tensor and an
				universal estimate for the effective shear modulus}},\ }\href
	{https://doi.org/10.1093/mnrasl/slaa173} {\bibfield  {journal} {\bibinfo
			{journal} {\mnras}\ }\textbf {\bibinfo {volume} {500}},\ \bibinfo {pages}
		{L17} (\bibinfo {year} {2021})},\ \Eprint {https://arxiv.org/abs/2010.08398}
	{arXiv:2010.08398 [astro-ph.HE]} \BibitemShut {NoStop}%
	\bibitem [{\citenamefont {{Chugunov}}(2022)}]{Chugunov22_elast_screen}%
	\BibitemOpen
	\bibfield  {author} {\bibinfo {author} {\bibfnamefont {A.~I.}\ \bibnamefont
			{{Chugunov}}},\ }\bibfield  {title} {\bibinfo {title} {{Neutron star crust in
				Voigt approximation II: general formula for electron screening correction for
				effective shear modulus}},\ }\href {https://doi.org/10.1093/mnras/stac2157} {\bibfield  {journal} {\bibinfo
			{journal} {\mnras}\ }\textbf {\bibinfo {volume} {517}},\ \bibinfo {pages}
		{4607} (\bibinfo {year} {2022})},
	\ \Eprint
	{https://arxiv.org/abs/2207.14649} {arXiv:2207.14649 [astro-ph.HE]}
	\BibitemShut {NoStop}%
	\bibitem [{\citenamefont {{Yakovlev}}(2023)}]{YakovkevDG23}%
	\BibitemOpen
	\bibfield  {author} {\bibinfo {author} {\bibfnamefont {D.~G.}\ \bibnamefont
			{{Yakovlev}}},\ }\bibfield  {title} {\bibinfo {title} {{Self-similarity
				relations for torsional oscillations of neutron stars}},\ }\href
	{https://doi.org/10.1093/mnras/stac2871} {\bibfield  {journal} {\bibinfo
			{journal} {\mnras}\ }\textbf {\bibinfo {volume} {518}},\ \bibinfo {pages}
		{1148} (\bibinfo {year} {2023})},\ \Eprint {https://arxiv.org/abs/2210.02931}
	{arXiv:2210.02931 [astro-ph.SR]} \BibitemShut {NoStop}%
	\bibitem [{\citenamefont {{Kozhberov}}\ and\ \citenamefont
		{{Yakovlev}}(2020)}]{KY20}%
	\BibitemOpen
	\bibfield  {author} {\bibinfo {author} {\bibfnamefont {A.~A.}\ \bibnamefont
			{{Kozhberov}}}\ and\ \bibinfo {author} {\bibfnamefont {D.~G.}\ \bibnamefont
			{{Yakovlev}}},\ }\bibfield  {title} {\bibinfo {title} {{Deformed crystals and
				torsional oscillations of neutron star crust}},\ }\bibfield  {journal}
	{\bibinfo  {journal} {\mnras}\ }\href
	{https://doi.org/10.1093/mnras/staa2715} {10.1093/mnras/staa2715} (\bibinfo
	{year} {2020}),\ \Eprint {https://arxiv.org/abs/2009.04952} {arXiv:2009.04952
		[astro-ph.HE]} \BibitemShut {NoStop}%
	\bibitem [{\citenamefont {{Passamonti}}\ and\ \citenamefont
		{{Pons}}(2016)}]{Passamonti_Pons_16}%
	\BibitemOpen
	\bibfield  {author} {\bibinfo {author} {\bibfnamefont {A.}~\bibnamefont
			{{Passamonti}}}\ and\ \bibinfo {author} {\bibfnamefont {J.~A.}\ \bibnamefont
			{{Pons}}},\ }\bibfield  {title} {\bibinfo {title} {{Quasi-periodic
				oscillations in superfluid, relativistic magnetars with nuclear pasta
				phases}},\ }\href {https://doi.org/10.1093/mnras/stw1880} {\bibfield
		{journal} {\bibinfo  {journal} {\mnras}\ }\textbf {\bibinfo {volume} {463}},\
		\bibinfo {pages} {1173} (\bibinfo {year} {2016})},\ \Eprint
	{https://arxiv.org/abs/1606.02132} {arXiv:1606.02132 [astro-ph.HE]}
	\BibitemShut {NoStop}%
	\bibitem [{\citenamefont {{Kobyakov}}\ and\ \citenamefont
		{{Pethick}}(2015)}]{kp15}%
	\BibitemOpen
	\bibfield  {author} {\bibinfo {author} {\bibfnamefont {D.}~\bibnamefont
			{{Kobyakov}}}\ and\ \bibinfo {author} {\bibfnamefont {C.~J.}\ \bibnamefont
			{{Pethick}}},\ }\bibfield  {title} {\bibinfo {title} {{Elastic properties of
				polycrystalline dense matter}},\ }\href
	{https://doi.org/10.1093/mnrasl/slv027} {\bibfield  {journal} {\bibinfo
			{journal} {\mnras}\ }\textbf {\bibinfo {volume} {449}},\ \bibinfo {pages}
		{L110} (\bibinfo {year} {2015})},\ \Eprint {https://arxiv.org/abs/1502.02461}
	{arXiv:1502.02461 [astro-ph.SR]} \BibitemShut {NoStop}%
	\bibitem [{\citenamefont {{Reuss}}(1929)}]{Reuss_29}%
	\BibitemOpen
	\bibfield  {author} {\bibinfo {author} {\bibfnamefont {A.}~\bibnamefont
			{{Reuss}}},\ }\bibfield  {title} {\bibinfo {title} {{Berechnung der
				Flie{\ss}grenze von Mischkristallen auf Grund der Plastizit{\"a}tsbedingung
				f{\"u}r Einkristalle .}},\ }\href {https://doi.org/10.1002/zamm.19290090104}
	{\bibfield  {journal} {\bibinfo  {journal} {Zeitschrift Angewandte Mathematik
				und Mechanik}\ }\textbf {\bibinfo {volume} {9}},\ \bibinfo {pages} {49}
		(\bibinfo {year} {1929})}\BibitemShut {NoStop}%
	\bibitem [{\citenamefont {{Hill}}(1952)}]{Hill_1952}%
	\BibitemOpen
	\bibfield  {author} {\bibinfo {author} {\bibfnamefont {R.}~\bibnamefont
			{{Hill}}},\ }\bibfield  {title} {\bibinfo {title} {{The Elastic Behaviour of
				a Crystalline Aggregate}},\ }\href
	{https://doi.org/10.1088/0370-1298/65/5/307} {\bibfield  {journal} {\bibinfo
			{journal} {Proceedings of the Physical Society A}\ }\textbf {\bibinfo
			{volume} {65}},\ \bibinfo {pages} {349} (\bibinfo {year} {1952})}\BibitemShut
	{NoStop}%
	\bibitem [{\citenamefont {{Kr{\"o}ner}}(1958)}]{Kroner58}%
	\BibitemOpen
	\bibfield  {author} {\bibinfo {author} {\bibfnamefont {E.}~\bibnamefont
			{{Kr{\"o}ner}}},\ }\bibfield  {title} {\bibinfo {title} {{Berechnung der
				elastischen Konstanten des Vielkristalls aus den Konstanten des
				Einkristalls}},\ }\href {https://doi.org/10.1007/BF01337948} {\bibfield
		{journal} {\bibinfo  {journal} {Zeitschrift fur Physik}\ }\textbf {\bibinfo
			{volume} {151}},\ \bibinfo {pages} {504} (\bibinfo {year}
		{1958})}\BibitemShut {NoStop}%
	\bibitem [{\citenamefont {{Eshelby}}(1961)}]{Eshelby61}%
	\BibitemOpen
	\bibfield  {author} {\bibinfo {author} {\bibfnamefont {J.~D.}\ \bibnamefont
			{{Eshelby}}},\ }\href@noop {} {\bibfield  {journal} {\bibinfo  {journal}
			{Prog.\ Solid Mech.}\ }\textbf {\bibinfo {volume} {2}},\ \bibinfo {pages}
		{87} (\bibinfo {year} {1961})}\BibitemShut {NoStop}%
	\bibitem [{\citenamefont {deWit}(2008)}]{deWit_sc_model}%
	\BibitemOpen
	\bibfield  {author} {\bibinfo {author} {\bibfnamefont {R.}~\bibnamefont
			{deWit}},\ }\bibfield  {title} {\bibinfo {title} {Elastic constants and
			thermal expansion averages of a nontextured polycrystal},\ }\href@noop {}
	{\bibfield  {journal} {\bibinfo  {journal} {Journal of mechanics of materials
				and structures}\ }\textbf {\bibinfo {volume} {3}},\ \bibinfo {pages} {195}
		(\bibinfo {year} {2008})}\BibitemShut {NoStop}%
	\bibitem [{\citenamefont {Wu}\ \emph {et~al.}(2023)\citenamefont {Wu},
		\citenamefont {Jia}, \citenamefont {Gou},\ and\ \citenamefont
		{Xu}}]{Diff_shape_inclus}%
	\BibitemOpen
	\bibfield  {author} {\bibinfo {author} {\bibfnamefont {Y.}~\bibnamefont
			{Wu}}, \bibinfo {author} {\bibfnamefont {M.}~\bibnamefont {Jia}}, \bibinfo
		{author} {\bibfnamefont {X.}~\bibnamefont {Gou}},\ and\ \bibinfo {author}
		{\bibfnamefont {W.}~\bibnamefont {Xu}},\ }\bibfield  {title} {\bibinfo
		{title} {Average eshelby tensor of an arbitrarily shaped inclusion from
			convexity to non-convexity: Effective elastic properties of composites},\
	}\href {https://doi.org/https://doi.org/10.1016/j.ijsolstr.2023.112183}
	{\bibfield  {journal} {\bibinfo  {journal} {International Journal of Solids
				and Structures}\ }\textbf {\bibinfo {volume} {269}},\ \bibinfo {pages}
		{112183} (\bibinfo {year} {2023})}\BibitemShut {NoStop}%
	\bibitem [{\citenamefont {{Hashin}}\ and\ \citenamefont
		{{Shtrikman}}(1962{\natexlab{a}})}]{HS62_a}%
	\BibitemOpen
	\bibfield  {author} {\bibinfo {author} {\bibfnamefont {Z.}~\bibnamefont
			{{Hashin}}}\ and\ \bibinfo {author} {\bibfnamefont {S.}~\bibnamefont
			{{Shtrikman}}},\ }\bibfield  {title} {\bibinfo {title} {{On some variational
				principles in anisotropic and nonhomogeneous elasticity}},\ }\href
	{https://doi.org/10.1016/0022-5096(62)90004-2} {\bibfield  {journal}
		{\bibinfo  {journal} {Journal of Mechanics Physics of Solids}\ }\textbf
		{\bibinfo {volume} {10}},\ \bibinfo {pages} {335} (\bibinfo {year}
		{1962}{\natexlab{a}})}\BibitemShut {NoStop}%
	\bibitem [{\citenamefont {{Hashin}}\ and\ \citenamefont
		{{Shtrikman}}(1962{\natexlab{b}})}]{HS62_b}%
	\BibitemOpen
	\bibfield  {author} {\bibinfo {author} {\bibfnamefont {Z.}~\bibnamefont
			{{Hashin}}}\ and\ \bibinfo {author} {\bibfnamefont {S.}~\bibnamefont
			{{Shtrikman}}},\ }\bibfield  {title} {\bibinfo {title} {{A variational
				approach to the theory of the elastic behaviour of polycrystals}},\ }\href
	{https://doi.org/10.1016/0022-5096(62)90005-4} {\bibfield  {journal}
		{\bibinfo  {journal} {Journal of Mechanics Physics of Solids}\ }\textbf
		{\bibinfo {volume} {10}},\ \bibinfo {pages} {343} (\bibinfo {year}
		{1962}{\natexlab{b}})}\BibitemShut {NoStop}%
	\bibitem [{\citenamefont {{Kube}}\ and\ \citenamefont
		{{Arguelles}}(2016)}]{Kube_Arguelles_16}%
	\BibitemOpen
	\bibfield  {author} {\bibinfo {author} {\bibfnamefont {C.~M.}\ \bibnamefont
			{{Kube}}}\ and\ \bibinfo {author} {\bibfnamefont {A.~P.}\ \bibnamefont
			{{Arguelles}}},\ }\bibfield  {title} {\bibinfo {title} {{Bounds and
				self-consistent estimates of the elastic constants of polycrystals}},\ }\href
	{https://doi.org/10.1016/j.cageo.2016.07.008} {\bibfield  {journal} {\bibinfo
			{journal} {Computers and Geosciences}\ }\textbf {\bibinfo {volume} {95}},\
		\bibinfo {pages} {118} (\bibinfo {year} {2016})}\BibitemShut {NoStop}%
	\bibitem [{\citenamefont {{Brown}}(2015)}]{Brown_15}%
	\BibitemOpen
	\bibfield  {author} {\bibinfo {author} {\bibfnamefont {J.~M.}\ \bibnamefont
			{{Brown}}},\ }\bibfield  {title} {\bibinfo {title} {{Determination of
				Hashin-Shtrikman bounds on the isotropic effective elastic moduli of
				polycrystals of any symmetry}},\ }\href
	{https://doi.org/10.1016/j.cageo.2015.03.009} {\bibfield  {journal} {\bibinfo
			{journal} {Computers and Geosciences}\ }\textbf {\bibinfo {volume} {80}},\
		\bibinfo {pages} {95} (\bibinfo {year} {2015})}\BibitemShut {NoStop}%
	\bibitem [{\citenamefont {{Wallace}}(1967)}]{Wallace67}%
	\BibitemOpen
	\bibfield  {author} {\bibinfo {author} {\bibfnamefont {D.~C.}\ \bibnamefont
			{{Wallace}}},\ }\bibfield  {title} {\bibinfo {title} {{Thermoelasticity of
				Stressed Materials and Comparison of Various Elastic Constants}},\ }\href
	{https://doi.org/10.1103/PhysRev.162.776} {\bibfield  {journal} {\bibinfo
			{journal} {Physical Review}\ }\textbf {\bibinfo {volume} {162}},\ \bibinfo
		{pages} {776} (\bibinfo {year} {1967})}\BibitemShut {NoStop}%
	\bibitem [{\citenamefont {{Blaschke}}(2017)}]{Blaschke17_elast}%
	\BibitemOpen
	\bibfield  {author} {\bibinfo {author} {\bibfnamefont {D.~N.}\ \bibnamefont
			{{Blaschke}}},\ }\bibfield  {title} {\bibinfo {title} {{Averaging of elastic
				constants for polycrystals}},\ }\href {https://doi.org/10.1063/1.4993443}
	{\bibfield  {journal} {\bibinfo  {journal} {Journal of Applied Physics}\
		}\textbf {\bibinfo {volume} {122}},\ \bibinfo {eid} {145110} (\bibinfo {year}
		{2017})},\ \Eprint {https://arxiv.org/abs/1706.07132} {arXiv:1706.07132
		[cond-mat.mtrl-sci]} \BibitemShut {NoStop}%
	\bibitem [{\citenamefont {Avellaneda}\ and\ \citenamefont
		{Milton}(1989)}]{Avellaneda_Milton89}%
	\BibitemOpen
	\bibfield  {author} {\bibinfo {author} {\bibfnamefont {M.}~\bibnamefont
			{Avellaneda}}\ and\ \bibinfo {author} {\bibfnamefont {G.~W.}\ \bibnamefont
			{Milton}},\ }\bibfield  {title} {\bibinfo {title} {Optimal bounds on the
			effective bulk modulus of polycrystals},\ }\href@noop {} {\bibfield
		{journal} {\bibinfo  {journal} {SIAM Journal on Applied Mathematics}\
		}\textbf {\bibinfo {volume} {49}},\ \bibinfo {pages} {824} (\bibinfo {year}
		{1989})}\BibitemShut {NoStop}%
	\bibitem [{\citenamefont {{Kr{\"o}ner}}(1977)}]{Kroner_77}%
	\BibitemOpen
	\bibfield  {author} {\bibinfo {author} {\bibfnamefont {E.}~\bibnamefont
			{{Kr{\"o}ner}}},\ }\bibfield  {title} {\bibinfo {title} {{Bounds for
				effective elastic moduli of disordered materials}},\ }\href
	{https://doi.org/10.1016/0022-5096(77)90009-6} {\bibfield  {journal}
		{\bibinfo  {journal} {Journal of Mechanics Physics of Solids}\ }\textbf
		{\bibinfo {volume} {25}},\ \bibinfo {pages} {137} (\bibinfo {year}
		{1977})}\BibitemShut {NoStop}%
	\bibitem [{\citenamefont {{Chamel}}\ \emph {et~al.}(2025)\citenamefont
		{{Chamel}}, \citenamefont {{Shchechilin}},\ and\ \citenamefont
		{{Chugunov}}}]{CSC25}%
	\BibitemOpen
	\bibfield  {author} {\bibinfo {author} {\bibfnamefont {N.}~\bibnamefont
			{{Chamel}}}, \bibinfo {author} {\bibfnamefont {N.~N.}\ \bibnamefont
			{{Shchechilin}}},\ and\ \bibinfo {author} {\bibfnamefont {A.~I.}\
			\bibnamefont {{Chugunov}}},\ }\bibfield  {title} {\bibinfo {title} {{Pressure
				and chemical potentials in the inner crust of a cold neutron star within
				Hartree-Fock and extended Thomas-Fermi methods}},\ }\href
	{https://doi.org/10.1103/PhysRevC.111.015805} {\bibfield  {journal} {\bibinfo
			{journal} {\prc}\ }\textbf {\bibinfo {volume} {111}},\ \bibinfo {eid}
		{015805} (\bibinfo {year} {2025})},\ \Eprint
	{https://arxiv.org/abs/2501.08075} {arXiv:2501.08075 [nucl-th]} \BibitemShut
	{NoStop}%
	\bibitem [{\citenamefont {{Zemlyakov}}\ and\ \citenamefont
		{{Chugunov}}(2023)}]{ZC23_shear_crust}%
	\BibitemOpen
	\bibfield  {author} {\bibinfo {author} {\bibfnamefont {N.~A.}\ \bibnamefont
			{{Zemlyakov}}}\ and\ \bibinfo {author} {\bibfnamefont {A.~I.}\ \bibnamefont
			{{Chugunov}}},\ }\bibfield  {title} {\bibinfo {title} {{Neutron star inner
				crust: reduction of shear modulus by nuclei finite size effect}},\ }\href
	{https://doi.org/10.1093/mnras/stac3377} {\bibfield  {journal} {\bibinfo
			{journal} {\mnras}\ }\textbf {\bibinfo {volume} {518}},\ \bibinfo {pages}
		{3813} (\bibinfo {year} {2023})},\ \Eprint {https://arxiv.org/abs/2209.05821}
	{arXiv:2209.05821 [astro-ph.HE]} \BibitemShut {NoStop}%
	\bibitem [{\citenamefont {{Xia}}\ \emph {et~al.}(2023)\citenamefont {{Xia}},
		\citenamefont {{Maruyama}}, \citenamefont {{Yasutake}}, \citenamefont
		{{Tatsumi}},\ and\ \citenamefont {{Zhang}}}]{Xia_ea23_Elast}%
	\BibitemOpen
	\bibfield  {author} {\bibinfo {author} {\bibfnamefont {C.-J.}\ \bibnamefont
			{{Xia}}}, \bibinfo {author} {\bibfnamefont {T.}~\bibnamefont {{Maruyama}}},
		\bibinfo {author} {\bibfnamefont {N.}~\bibnamefont {{Yasutake}}}, \bibinfo
		{author} {\bibfnamefont {T.}~\bibnamefont {{Tatsumi}}},\ and\ \bibinfo
		{author} {\bibfnamefont {Y.-X.}\ \bibnamefont {{Zhang}}},\ }\bibfield
	{title} {\bibinfo {title} {{Elastic properties of nuclear pasta in a fully
				three-dimensional geometry}},\ }\href
	{https://doi.org/10.1016/j.physletb.2023.137769} {\bibfield  {journal}
		{\bibinfo  {journal} {Physics Letters B}\ }\textbf {\bibinfo {volume}
			{839}},\ \bibinfo {eid} {137769} (\bibinfo {year} {2023})},\ \Eprint
	{https://arxiv.org/abs/2209.13310} {arXiv:2209.13310 [nucl-th]} \BibitemShut
	{NoStop}%
	\bibitem [{\citenamefont {{Eshelby}}(1957)}]{Eshelby57}%
	\BibitemOpen
	\bibfield  {author} {\bibinfo {author} {\bibfnamefont {J.~D.}\ \bibnamefont
			{{Eshelby}}},\ }\bibfield  {title} {\bibinfo {title} {{The Determination of
				the Elastic Field of an Ellipsoidal Inclusion, and Related Problems}},\
	}\href {https://doi.org/10.1098/rspa.1957.0133} {\bibfield  {journal}
		{\bibinfo  {journal} {Proceedings of the Royal Society of London Series A}\
		}\textbf {\bibinfo {volume} {241}},\ \bibinfo {pages} {376} (\bibinfo {year}
		{1957})}\BibitemShut {NoStop}%
	\bibitem [{\citenamefont {{Jancovici}}(1962)}]{Jancovici62}%
	\BibitemOpen
	\bibfield  {author} {\bibinfo {author} {\bibfnamefont {B.}~\bibnamefont
			{{Jancovici}}},\ }\bibfield  {title} {\bibinfo {title} {{On the relativistic
				degenerate electron gas}},\ }\href {https://doi.org/10.1007/BF02731458}
	{\bibfield  {journal} {\bibinfo  {journal} {Il Nuovo Cimento}\ }\textbf
		{\bibinfo {volume} {25}},\ \bibinfo {pages} {428} (\bibinfo {year}
		{1962})}\BibitemShut {NoStop}%
	\bibitem [{\citenamefont {{Fortov}}\ \emph {et~al.}(2005)\citenamefont
		{{Fortov}}, \citenamefont {{Ivlev}}, \citenamefont {{Khrapak}}, \citenamefont
		{{Khrapak}},\ and\ \citenamefont {{Morfill}}}]{Fortov_ea05}%
	\BibitemOpen
	\bibfield  {author} {\bibinfo {author} {\bibfnamefont {V.~E.}\ \bibnamefont
			{{Fortov}}}, \bibinfo {author} {\bibfnamefont {A.~V.}\ \bibnamefont
			{{Ivlev}}}, \bibinfo {author} {\bibfnamefont {S.~A.}\ \bibnamefont
			{{Khrapak}}}, \bibinfo {author} {\bibfnamefont {A.~G.}\ \bibnamefont
			{{Khrapak}}},\ and\ \bibinfo {author} {\bibfnamefont {G.~E.}\ \bibnamefont
			{{Morfill}}},\ }\bibfield  {title} {\bibinfo {title} {{Complex (dusty)
				plasmas: Current status, open issues, perspectives}},\ }\href
	{https://doi.org/10.1016/j.physrep.2005.08.007} {\bibfield  {journal}
		{\bibinfo  {journal} {\physrep}\ }\textbf {\bibinfo {volume} {421}},\
		\bibinfo {pages} {1} (\bibinfo {year} {2005})}\BibitemShut {NoStop}%
	\bibitem [{\citenamefont {{Donk{\'o}}}\ \emph {et~al.}(2008)\citenamefont
		{{Donk{\'o}}}, \citenamefont {{Kalman}},\ and\ \citenamefont
		{{Hartmann}}}]{Donko_ea08_Review}%
	\BibitemOpen
	\bibfield  {author} {\bibinfo {author} {\bibfnamefont {Z.}~\bibnamefont
			{{Donk{\'o}}}}, \bibinfo {author} {\bibfnamefont {G.~J.}\ \bibnamefont
			{{Kalman}}},\ and\ \bibinfo {author} {\bibfnamefont {P.}~\bibnamefont
			{{Hartmann}}},\ }\bibfield  {title} {\bibinfo {title} {{Dynamical correlations and collective excitations of Yukawa liquids}},\
	}\href {https://doi.org/10.1088/0953-8984/20/41/413101} {\bibfield  {journal}
		{\bibinfo  {journal} {Journal of Physics Condensed Matter}\ }\textbf
		{\bibinfo {volume} {20}},\ \bibinfo {eid} {413101} (\bibinfo {year}
		{2008})},\ \Eprint {https://arxiv.org/abs/0808.1963} {arXiv:0808.1963
		[physics.plasm-ph]} \BibitemShut {NoStop}%
	\bibitem [{\citenamefont {{Klumov}}(2010)}]{Klumov10_Melting}%
	\BibitemOpen
	\bibfield  {author} {\bibinfo {author} {\bibfnamefont {B.~A.}\ \bibnamefont
			{{Klumov}}},\ }\bibfield  {title} {\bibinfo {title} {{On
				melting criteria for complex plasma}},\ }\href
	{https://doi.org/10.3367/UFNe.0180.201010e.1095} {\bibfield  {journal}
		{\bibinfo  {journal} {Physics Uspekhi}\ }\textbf {\bibinfo {volume} {53}},\
		\bibinfo {pages} {1053} (\bibinfo {year} {2010})}\BibitemShut {NoStop}%
	\bibitem [{\citenamefont {{Khrapak}}(2016)}]{KhrapakSA16_Yuk_thermod}%
	\BibitemOpen
	\bibfield  {author} {\bibinfo {author} {\bibfnamefont {S.~A.}\ \bibnamefont
			{{Khrapak}}},\ }\bibfield  {title} {\bibinfo {title} {{Thermodynamics of
				Yukawa systems and sound velocity in dusty plasmas}},\ }\href
	{https://doi.org/10.1088/0741-3335/58/1/014022} {\bibfield  {journal}
		{\bibinfo  {journal} {Plasma Physics and Controlled Fusion}\ }\textbf
		{\bibinfo {volume} {58}},\ \bibinfo {eid} {014022} (\bibinfo {year}
		{2016})}\BibitemShut {NoStop}%
	\bibitem [{\citenamefont {{Khrapak}}\ and\ \citenamefont
		{{Klumov}}(2018)}]{Khrapak_Klumov18_2D_Yukawa_elast}%
	\BibitemOpen
	\bibfield  {author} {\bibinfo {author} {\bibfnamefont {S.}~\bibnamefont
			{{Khrapak}}}\ and\ \bibinfo {author} {\bibfnamefont {B.}~\bibnamefont
			{{Klumov}}},\ }\bibfield  {title} {\bibinfo {title} {{High-frequency elastic
				moduli of two-dimensional Yukawa fluids and solids}},\ }\href
	{https://doi.org/10.1063/1.5025396} {\bibfield  {journal} {\bibinfo
			{journal} {Physics of Plasmas}\ }\textbf {\bibinfo {volume} {25}},\ \bibinfo
		{eid} {033706} (\bibinfo {year} {2018})},\ \Eprint
	{https://arxiv.org/abs/1803.05295} {arXiv:1803.05295 [physics.plasm-ph]}
	\BibitemShut {NoStop}%
	\bibitem [{\citenamefont {Lu}\ \emph {et~al.}(2022)\citenamefont {Lu},
		\citenamefont {Huang},\ and\ \citenamefont
		{Feng}}]{Lu_ea22_2D_Yukawa_ShearSoftening}%
	\BibitemOpen
	\bibfield  {author} {\bibinfo {author} {\bibfnamefont {S.}~\bibnamefont
			{Lu}}, \bibinfo {author} {\bibfnamefont {D.}~\bibnamefont {Huang}},\ and\
		\bibinfo {author} {\bibfnamefont {Y.}~\bibnamefont {Feng}},\ }\bibfield
	{title} {\bibinfo {title} {Shear softening and hardening of a two-dimensional
			yukawa solid},\ }\href {https://doi.org/10.1103/PhysRevE.105.035203}
	{\bibfield  {journal} {\bibinfo  {journal} {Phys. Rev. E}\ }\textbf {\bibinfo
			{volume} {105}},\ \bibinfo {pages} {035203} (\bibinfo {year}
		{2022})}\BibitemShut {NoStop}%
	\bibitem [{\citenamefont {Beckers}\ \emph {et~al.}(2023)\citenamefont
		{Beckers}, \citenamefont {Berndt}, \citenamefont {Block}, \citenamefont
		{Bonitz}, \citenamefont {Bruggeman}, \citenamefont {CouГ«del}, \citenamefont
		{Delzanno}, \citenamefont {Feng}, \citenamefont {Gopalakrishnan},
		\citenamefont {Greiner}, \citenamefont {Hartmann}, \citenamefont {HorГЎnyi},
		\citenamefont {Kersten}, \citenamefont {Knapek}, \citenamefont {Konopka},
		\citenamefont {Kortshagen}, \citenamefont {Kostadinova}, \citenamefont
		{KovaДЌeviД‡}, \citenamefont {Krasheninnikov}, \citenamefont {Mann},
		\citenamefont {Mariotti}, \citenamefont {Matthews}, \citenamefont {Melzer},
		\citenamefont {Mikikian}, \citenamefont {Nosenko}, \citenamefont {Pustylnik},
		\citenamefont {Ratynskaia}, \citenamefont {Sankaran}, \citenamefont
		{Schneider}, \citenamefont {Thimsen}, \citenamefont {Thomas}, \citenamefont
		{Thomas}, \citenamefont {Tolias},\ and\ \citenamefont {van~de
			Kerkhof}}]{Beckers_ea23_DustyPlasma_Perspectives23}%
	\BibitemOpen
	\bibfield  {author} {\bibinfo {author} {\bibfnamefont {J.}~\bibnamefont
			{Beckers}}, \bibinfo {author} {\bibfnamefont {J.}~\bibnamefont {Berndt}},
		\bibinfo {author} {\bibfnamefont {D.}~\bibnamefont {Block}}, \bibinfo
		{author} {\bibfnamefont {M.}~\bibnamefont {Bonitz}}, \bibinfo {author}
		{\bibfnamefont {P.~J.}\ \bibnamefont {Bruggeman}}, \bibinfo {author}
		{\bibfnamefont {L.}~\bibnamefont {Couedel}}, \bibinfo {author}
		{\bibfnamefont {G.~L.}\ \bibnamefont {Delzanno}}, \bibinfo {author}
		{\bibfnamefont {Y.}~\bibnamefont {Feng}}, \bibinfo {author} {\bibfnamefont
			{R.}~\bibnamefont {Gopalakrishnan}}, \bibinfo {author} {\bibfnamefont
			{F.}~\bibnamefont {Greiner}}, \bibinfo {author} {\bibfnamefont
			{P.}~\bibnamefont {Hartmann}}, \bibinfo {author} {\bibfnamefont
			{M.}~\bibnamefont {Horanyi}}, \bibinfo {author} {\bibfnamefont
			{H.}~\bibnamefont {Kersten}}, \bibinfo {author} {\bibfnamefont {C.~A.}\
			\bibnamefont {Knapek}}, \bibinfo {author} {\bibfnamefont {U.}~\bibnamefont
			{Konopka}}, \bibinfo {author} {\bibfnamefont {U.}~\bibnamefont {Kortshagen}},
		\bibinfo {author} {\bibfnamefont {E.~G.}\ \bibnamefont {Kostadinova}},
		\bibinfo {author} {\bibfnamefont {E.}~\bibnamefont {Kovacevic}}, \bibinfo
		{author} {\bibfnamefont {S.~I.}\ \bibnamefont {Krasheninnikov}}, \bibinfo
		{author} {\bibfnamefont {I.}~\bibnamefont {Mann}}, \bibinfo {author}
		{\bibfnamefont {D.}~\bibnamefont {Mariotti}}, \bibinfo {author}
		{\bibfnamefont {L.~S.}\ \bibnamefont {Matthews}}, \bibinfo {author}
		{\bibfnamefont {A.}~\bibnamefont {Melzer}}, \bibinfo {author} {\bibfnamefont
			{M.}~\bibnamefont {Mikikian}}, \bibinfo {author} {\bibfnamefont
			{V.}~\bibnamefont {Nosenko}}, \bibinfo {author} {\bibfnamefont {M.~Y.}\
			\bibnamefont {Pustylnik}}, \bibinfo {author} {\bibfnamefont {S.}~\bibnamefont
			{Ratynskaia}}, \bibinfo {author} {\bibfnamefont {R.~M.}\ \bibnamefont
			{Sankaran}}, \bibinfo {author} {\bibfnamefont {V.}~\bibnamefont {Schneider}},
		\bibinfo {author} {\bibfnamefont {E.~J.}\ \bibnamefont {Thimsen}}, \bibinfo
		{author} {\bibfnamefont {E.}~\bibnamefont {Thomas}}, \bibinfo {author}
		{\bibfnamefont {H.~M.}\ \bibnamefont {Thomas}}, \bibinfo {author}
		{\bibfnamefont {P.}~\bibnamefont {Tolias}},\ and\ \bibinfo {author}
		{\bibfnamefont {M.}~\bibnamefont {van~de Kerkhof}},\ }\bibfield  {title}
	{\bibinfo {title} {Physics and applications of dusty plasmas: The
			perspectives 2023},\ }\href {https://doi.org/10.1063/5.0168088} {\bibfield
		{journal} {\bibinfo  {journal} {Physics of Plasmas}\ }\textbf {\bibinfo
			{volume} {30}},\ \bibinfo {pages} {120601} (\bibinfo {year}
		{2023})}\BibitemShut {NoStop}%
	\bibitem [{\citenamefont {{Zhukhovitskii}}\ and\ \citenamefont
		{{Perevoshchikov}}(2024)}]{Zhukhovitskii_Perevoschikov24}%
	\BibitemOpen
	\bibfield  {author} {\bibinfo {author} {\bibfnamefont {D.}~\bibnamefont
			{{Zhukhovitskii}}}\ and\ \bibinfo {author} {\bibfnamefont {E.}~\bibnamefont
			{{Perevoshchikov}}},\ }\bibfield  {title} {\bibinfo {title} {{Structural
				Transition in Strongly Coupled Coulomb Clusters}},\ }\href
	{https://doi.org/10.1134/S0018151X25700142} {\bibfield  {journal} {\bibinfo
			{journal} {High Temperature}\ }\textbf {\bibinfo {volume} {62}},\ \bibinfo
		{pages} {421} (\bibinfo {year} {2024})}\BibitemShut {NoStop}%
	\bibitem [{\citenamefont {{Lipaev}}\ \emph {et~al.}(2025)\citenamefont
		{{Lipaev}}, \citenamefont {{Naumkin}}, \citenamefont {{Khrapak}},
		\citenamefont {{Usachev}}, \citenamefont {{Petrov}}, \citenamefont {{Thoma}},
		\citenamefont {{Kretschmer}}, \citenamefont {{Du}}, \citenamefont
		{{Kononenko}},\ and\ \citenamefont {{Zobnin}}}]{Lipaev_ea25_WaveDispersion}%
	\BibitemOpen
	\bibfield  {author} {\bibinfo {author} {\bibfnamefont {A.~M.}\ \bibnamefont
			{{Lipaev}}}, \bibinfo {author} {\bibfnamefont {V.~N.}\ \bibnamefont
			{{Naumkin}}}, \bibinfo {author} {\bibfnamefont {S.~A.}\ \bibnamefont
			{{Khrapak}}}, \bibinfo {author} {\bibfnamefont {A.~D.}\ \bibnamefont
			{{Usachev}}}, \bibinfo {author} {\bibfnamefont {O.~F.}\ \bibnamefont
			{{Petrov}}}, \bibinfo {author} {\bibfnamefont {M.~H.}\ \bibnamefont
			{{Thoma}}}, \bibinfo {author} {\bibfnamefont {M.}~\bibnamefont
			{{Kretschmer}}}, \bibinfo {author} {\bibfnamefont {C.-R.}\ \bibnamefont
			{{Du}}}, \bibinfo {author} {\bibfnamefont {O.~D.}\ \bibnamefont
			{{Kononenko}}},\ and\ \bibinfo {author} {\bibfnamefont {A.~V.}\ \bibnamefont
			{{Zobnin}}},\ }\bibfield  {title} {\bibinfo {title} {{Wave dispersion in a
				three-dimensional complex plasma solid under microgravity conditions}},\
	}\href {https://doi.org/10.1103/PhysRevE.111.015209} {\bibfield  {journal}
		{\bibinfo  {journal} {\pre}\ }\textbf {\bibinfo {volume} {111}},\ \bibinfo
		{eid} {015209} (\bibinfo {year} {2025})}\BibitemShut {NoStop}%
	\bibitem [{\citenamefont {{Baiko}}\ and\ \citenamefont
		{{Chugunov}}(2022)}]{BC21}%
	\BibitemOpen
	\bibfield  {author} {\bibinfo {author} {\bibfnamefont {D.~A.}\ \bibnamefont
			{{Baiko}}}\ and\ \bibinfo {author} {\bibfnamefont {A.~I.}\ \bibnamefont
			{{Chugunov}}},\ }\bibfield  {title} {\bibinfo {title} {{Ab initio
				thermodynamics of one-component plasma for astrophysics of white dwarfs and
				neutron stars}},\ }\href {https://doi.org/10.1093/mnras/stab3613} {\bibfield
		{journal} {\bibinfo  {journal} {\mnras}\ }\textbf {\bibinfo {volume} {510}},\
		\bibinfo {pages} {2628} (\bibinfo {year} {2022})},\ \Eprint
	{https://arxiv.org/abs/2112.04822} {arXiv:2112.04822 [astro-ph.HE]}
	\BibitemShut {NoStop}%
	\bibitem [{\citenamefont {{Chamel}}\ and\ \citenamefont
		{{Fantina}}(2016)}]{cf16_mix}%
	\BibitemOpen
	\bibfield  {author} {\bibinfo {author} {\bibfnamefont {N.}~\bibnamefont
			{{Chamel}}}\ and\ \bibinfo {author} {\bibfnamefont {A.~F.}\ \bibnamefont
			{{Fantina}}},\ }\bibfield  {title} {\bibinfo {title} {{Binary and ternary
				ionic compounds in the outer crust of a cold nonaccreting neutron star}},\
	}\href {https://doi.org/10.1103/PhysRevC.94.065802} {\bibfield  {journal}
		{\bibinfo  {journal} {\prc}\ }\textbf {\bibinfo {volume} {94}},\ \bibinfo
		{eid} {065802} (\bibinfo {year} {2016})}\BibitemShut {NoStop}%
	\bibitem [{\citenamefont {{Kozhberov}}\ and\ \citenamefont
		{{Baiko}}(2012)}]{kb12}%
	\BibitemOpen
	\bibfield  {author} {\bibinfo {author} {\bibfnamefont {A.~A.}\ \bibnamefont
			{{Kozhberov}}}\ and\ \bibinfo {author} {\bibfnamefont {D.~A.}\ \bibnamefont
			{{Baiko}}},\ }\bibfield  {title} {\bibinfo {title} {{Physical Features of
				Binary Coulomb Crystals: Madelung Energy, Collective Modes and Phonon Heat
				Capacity}},\ }\href {https://doi.org/10.1002/ctpp.201100091} {\bibfield
		{journal} {\bibinfo  {journal} {Contributions to Plasma Physics}\ }\textbf
		{\bibinfo {volume} {52}},\ \bibinfo {pages} {153} (\bibinfo {year}
		{2012})}\BibitemShut {NoStop}%
	\bibitem [{\citenamefont {{Salpeter}}(1954)}]{Salpeter54}%
	\BibitemOpen
	\bibfield  {author} {\bibinfo {author} {\bibfnamefont {E.~E.}\ \bibnamefont
			{{Salpeter}}},\ }\bibfield  {title} {\bibinfo {title} {{Electrons Screening
				and Thermonuclear Reactions}},\ }\href {https://doi.org/10.1071/PH540373}
	{\bibfield  {journal} {\bibinfo  {journal} {Australian Journal of Physics}\
		}\textbf {\bibinfo {volume} {7}},\ \bibinfo {pages} {373} (\bibinfo {year}
		{1954})}\BibitemShut {NoStop}%
	\bibitem [{\citenamefont {{Hansen}}\ and\ \citenamefont
		{{Vieillefosse}}(1976)}]{HV76}%
	\BibitemOpen
	\bibfield  {author} {\bibinfo {author} {\bibfnamefont {J.~P.}\ \bibnamefont
			{{Hansen}}}\ and\ \bibinfo {author} {\bibfnamefont {P.}~\bibnamefont
			{{Vieillefosse}}},\ }\bibfield  {title} {\bibinfo {title} {{Equation of state
				of the classical two-component plasma}},\ }\href
	{https://doi.org/10.1103/PhysRevLett.37.391} {\bibfield  {journal} {\bibinfo
			{journal} {\prl}\ }\textbf {\bibinfo {volume} {37}},\ \bibinfo {pages} {391}
		(\bibinfo {year} {1976})}\BibitemShut {NoStop}%
	\bibitem [{\citenamefont {{Dederichs}}\ and\ \citenamefont
		{{Zeller}}(1973)}]{Dederichs_Zeller_73}%
	\BibitemOpen
	\bibfield  {author} {\bibinfo {author} {\bibfnamefont {P.~H.}\ \bibnamefont
			{{Dederichs}}}\ and\ \bibinfo {author} {\bibfnamefont {R.}~\bibnamefont
			{{Zeller}}},\ }\bibfield  {title} {\bibinfo {title} {{Variational treatment
				of the elastic constants of disordered materials}},\ }\href
	{https://doi.org/10.1007/BF01392841} {\bibfield  {journal} {\bibinfo
			{journal} {Zeitschrift fur Physik}\ }\textbf {\bibinfo {volume} {259}},\
		\bibinfo {pages} {103} (\bibinfo {year} {1973})}\BibitemShut {NoStop}%
	\bibitem [{\citenamefont {Gairola}\ and\ \citenamefont
		{Kr{\"o}ner}(1981)}]{Gairola_Kroner_81}%
	\BibitemOpen
	\bibfield  {author} {\bibinfo {author} {\bibfnamefont {B.}~\bibnamefont
			{Gairola}}\ and\ \bibinfo {author} {\bibfnamefont {E.}~\bibnamefont
			{Kr{\"o}ner}},\ }\bibfield  {title} {\bibinfo {title} {A simple formula for
			calculating the bounds and the self-consistent value of the shear modulus of
			a polycrystalline aggregate of cubic crystals},\ }\href
	{https://doi.org/https://doi.org/10.1016/0020-7225(81)90120-8} {\bibfield
		{journal} {\bibinfo  {journal} {International Journal of Engineering
				Science}\ }\textbf {\bibinfo {volume} {19}},\ \bibinfo {pages} {865}
		(\bibinfo {year} {1981})}\BibitemShut {NoStop}%
	\bibitem [{\citenamefont {{Kube}}\ and\ \citenamefont {{de
				Jong}}(2016)}]{Kube_Jong_16}%
	\BibitemOpen
	\bibfield  {author} {\bibinfo {author} {\bibfnamefont {C.~M.}\ \bibnamefont
			{{Kube}}}\ and\ \bibinfo {author} {\bibfnamefont {M.}~\bibnamefont {{de
					Jong}}},\ }\bibfield  {title} {\bibinfo {title} {{Elastic constants of
				polycrystals with generally anisotropic crystals}},\ }\href
	{https://doi.org/10.1063/1.4965867} {\bibfield  {journal} {\bibinfo
			{journal} {Journal of Applied Physics}\ }\textbf {\bibinfo {volume} {120}},\
		\bibinfo {eid} {165105} (\bibinfo {year} {2016})}\BibitemShut {NoStop}%
	\bibitem [{\citenamefont {{Berryman}}(2005)}]{Berryman_05}%
	\BibitemOpen
	\bibfield  {author} {\bibinfo {author} {\bibfnamefont {J.~G.}\ \bibnamefont
			{{Berryman}}},\ }\bibfield  {title} {\bibinfo {title} {{Bounds and
				self-consistent estimates for elastic constants of random polycrystals with
				hexagonal, trigonal, and tetragonal symmetries}},\ }\href
	{https://doi.org/10.1016/j.jmps.2005.05.004} {\bibfield  {journal} {\bibinfo
			{journal} {Journal of Mechanics Physics of Solids}\ }\textbf {\bibinfo
			{volume} {53}},\ \bibinfo {pages} {2141} (\bibinfo {year}
		{2005})}\BibitemShut {NoStop}%
\end{thebibliography}
%

\end{document}